\documentclass[prd,onecolumn,aps,nofootinbib,eqsecnum,notitlepage,nobibnotes,superscriptaddress,preprintnumbers]{revtex4-1}

\usepackage{here}
\usepackage{graphicx}
\usepackage{amsmath,amsthm,amssymb}
\usepackage{bm}
\usepackage{color}

\usepackage{amsfonts}
\usepackage{dcolumn}
\usepackage{hyperref}
\allowdisplaybreaks[1]
\usepackage{ulem}

\begin{document}
\newcommand{\newc}{\newcommand}

\newcommand{\rk}{\textcolor{red}}
\newc{\ben}{\begin{eqnarray}}
\newc{\een}{\end{eqnarray}}
\newc{\be}{\begin{equation}}
\newc{\ee}{\end{equation}}
\newc{\ba}{\begin{eqnarray}}
\newc{\ea}{\end{eqnarray}}
\newc{\D}{\partial}
\newc{\rH}{{\rm H}}
\newc{\rd}{{\rm d}}

\newcommand{\mm}[1]{\textcolor{red}{#1}}
\newcommand{\mmc}[1]{\textcolor{green}{[MM:~#1]}}

\title{Spontaneous scalarization of charged stars}

\author{Masato Minamitsuji}
\affiliation{
Centro de Astrof\'{\i}sica e Gravita\c c\~ao  - CENTRA, Departamento de F\'{\i}sica, Instituto Superior T\'ecnico - IST, Universidade de Lisboa - UL, Av. Rovisco Pais 1, 1049-001 Lisboa, Portugal}

\author{Shinji Tsujikawa}
\affiliation{
Department of Physics, Waseda University, 3-4-1 Okubo, Shinjuku, Tokyo 169-8555, Japan}

\begin{abstract}

We study static and spherically symmetric charged stars with a nontrivial profile of the scalar field 
$\phi$ in Einstein-Maxwell-scalar theories. The scalar field is coupled to a $U(1)$ gauge field 
$A_{\mu}$ with the form $-\alpha(\phi)F_{\mu \nu}F^{\mu \nu}/4$, where 
$F_{\mu \nu}=\partial_{\mu}A_{\nu}-\partial_{\nu} A_{\mu}$ is the field strength tensor.
Analogous to the case of charged black holes, 
we show that this type of interaction can induce spontaneous scalarization of charged stars under the conditions $({\rm d}\alpha/{\rm d}\phi) (0)=0$ and $({\rm d}^2\alpha/{\rm d}\phi^2) (0)>0$. For the coupling $\alpha (\phi)=\exp (-\beta \phi^2/M_{\rm pl}^2)$, where $\beta~(<0)$ is a coupling constant and 
$M_{\rm pl}$ is a reduced Planck mass, there is a branch of charged star solutions with a nontrivial profile of $\phi$ approaching $0$ toward spatial infinity, besides a branch of general relativistic solutions 
with a vanishing scalar field, i.e., solutions in the Einstein-Maxwell model. 
As the ratio $\rho_c/\rho_m$ between charge density $\rho_c$ and matter density $\rho_m$ 
increases toward its maximum value, the mass $M$ of charged stars in general relativity 
tends to be enhanced due to the increase of repulsive Coulomb force against gravity. 
In this regime, the appearance of nontrivial branches induced by negative $\beta$ 
of order $-1$ effectively reduces the Coulomb force for a wide range of central matter densities, 
leading to charged stars with smaller masses and radii in comparison to those in the general 
relativistic branch. 
Our analysis indicates that spontaneous scalarization of stars can be induced not only by the coupling to curvature invariants but also by the scalar-gauge coupling in Einstein gravity.

\end{abstract}

\date{\today}

%\pacs{04.50.Kd, 95.36.+x, 98.80.-k}

\maketitle

%%%%%%%%%%%%%%%%%%%%%%%%%%%%%%%%%%%%%%%%%%
\section{Introduction}
\label{introsec}
%%%%%%%%%%%%%%%%%%%%%%%%%%%%%%%%%%%%%%%%%%

The detection of gravitational waves emitted from the binary systems of 
black holes \cite{Abbott2016} and neutron stars \cite{GW170817} 
opened up a new window for probing physics in the strong gravity regime. 
The upcoming high-precision observational data
will allow us to test the accuracy of General Relativity (GR) and the possible 
deviation from it \cite{Berti,Barack,Berti:2018cxi}. 
In this sense, it is important to scrutinize observational signatures of 
theories beyond GR in the strong gravitational background. 
In scalar-tensor or vector-tensor theories, for example, the scalar or 
vector field coupled to gravity can give rise to hairy solutions of 
black holes and neutron stars which are distinguished from 
no-hair 
solutions in GR \cite{Damour:1993hw,Damour2,Kanti:1995vq,Alexeev:1996vs,Cooney:2009rr,Arapoglu:2010rz,Rinaldi,Minami13,Yazadjiev1,Soti1,Soti2,Babi17,Koba14,Babi14,Yazadjiev2,Babi16,Tasinato2,Minamitsuji,GPBH,GPBH2,Fan,Cisterna,Babichev17,KMT17,Kase:2018owh,Kase:2019dqc,BenAchour:2018dap,Kobayashi:2018xvr,BenAchour:2018dap,Motohashi:2019sen,Minamitsuji:2019tet}.

In the context of neutron stars, Damour and Esposito-Farese \cite{Damour:1993hw,Damour2} 
have shown that a phenomenon dubbed spontaneous scalarization can occur for a scalar field $\phi$ 
coupled to a Ricci scalar $R$ of the form $F(\phi)R$. 
If the coupling $F(\phi)$ satisfies 
$({\rm d}F/{\rm d})\phi (0)=0$ and $({\rm d}^2F/{\rm d}\phi^2)(0)>0$, 
there is a possibility of having a static and spherically symmetric solution with a scalar hair, 
besides the GR branch with $\phi=0$.
For example, the nonminimal coupling 
$F(\phi)=e^{-\tilde{\beta} \phi^2/(2M_{\rm pl}^2)}$ with a negative 
constant ${\tilde \beta}$ in the range $\tilde \beta \lesssim -4.35$
can induce spontaneous scalarization, where 
the upper bound of $\tilde{\beta}$ is insensitive to the change of the 
equation of state of stars \cite{Harada:1998ge,Novak:1998rk,Silva:2014fca}.
Moreover, the stability analysis against both even- and odd-parity 
perturbations shows that the scalarized solution is 
stable for $F(\phi)>0$ \cite{Kase:2020qvz}.
The observational signatures of scalarized 
solutions \cite{Sotani1,Freire:2012mg,Sotani2,Pappas:2015npa} 
and extensions to rotating solutions \cite{Sotani:2012eb,Doneva:2013qva,Silva:2014fca,Doneva:2014faa,Pani:2014jra,Doneva:2014uma,Minamitsuji:2016hkk} have been extensively 
studied in the literature. 
The couplings of vector or tensor fields with curvature invariants 
also give rise to similar phenomena 
such as spontaneous vectorization, tensorization, and 
spinorization of relativistic stars~\cite{Annulli:2019fzq,Ramazanoglu:2017xbl,Ramazanoglu:2019gbz,Kase:2020yhw,Minamitsuji:2020pak,Ramazanoglu:2018hwk,Minamitsuji:2020hpl}.

Spontaneous scalarization can also occur for static and spherically 
symmetric black holes in the presence of
a Gauss-Bonnet term $R_{\rm GB}^2$ coupled to a scalar field $\phi$ of 
the form $\xi (\phi) R_{\rm GB}^2$ 
(see e.g., \cite{Doneva:2017bvd,Doneva:2017bvd,Silva:2017uqg,Silva:2018qhn,Silva:2017uqg,Antoniou:2017acq,Antoniou:2017hxj,Minamitsuji:2018xde}).
As in nonmininally coupled scalar-tensor theories mentioned above, 
this is regarded as a curvature-induced scalarization of compact objects 
triggering tachyonic instabilities of the GR branch toward hairy solutions in high 
curvature/density backgrounds. 
For the coupling $\xi (\phi)=\eta \phi^2$ with $\eta<0$,  
there is also a phenomenon of spin-induced black hole scalarizations
where rapidly rotating Kerr solutions in GR can exhibit tachyonic 
instabilities toward stationary and axisymmetric solutions 
with scalar hair \cite{Cunha,Dima:2020yac,Hod:2020jjy,Herdeiro:2020wei,Berti:2020kgk,Doneva:2021dqn}.
Scalarization of rotating black holes can also occur
in dynamical Chern-Simons theories \cite{Gao:2018acg,Myung:2020etf,Doneva:2021dcc}.

The occurrence of spontaneous scalarization does not necessarily 
require the existence of couplings to curvature invariants, but 
the scalar field $\phi$ coupled to a gauge field $A_{\mu}$
of  the form $-\alpha(\phi) F_{\mu \nu}F^{\mu \nu}/4$ 
with $F_{\mu\nu}=\partial_\mu A_\nu-\partial_\nu A_\mu$
can induce scalarization of black holes even 
in Einstein gravity \cite{Stefanov,Herdeiro1,Myung:2018vug,Herdeiro2,Brihaye:2019kvj,Myung:2019oua,Herdeiro3,Ikeda:2019okp,Hod:2020ljo}.
Hairy black hole solutions in this 
Einstein-Maxwell-scalar theory were originally studied by 
Gibbons and Maeda \cite{Gibbons:1987ps} 
and Garfinkle {\it et al.} \cite{Garfinkle:1990qj}
in higher-dimensional theories (see also 
Refs.~\cite{Mignemi:1992nt,Torii:1996yi}).
In string theory, for example, there is a dilaton field $\phi$ 
coupled to the gauge field of 
the form $\alpha(\phi)=e^{-\phi}$ in the Einstein frame. 
On the other hand, if we consider a coupling $\alpha(\phi)$ 
satisfying the conditions 
$({\rm d}\alpha/{\rm d}\phi) (0)=0$ and 
$({\rm d}^2\alpha/{\rm d}\phi^2)(0)>0$,
e.g., $\alpha (\phi)=\exp (-\beta \phi^2/M_{\rm pl}^2)$, 
there exist charge-induced scalarized black hole solutions
besides the GR branch with a vanishing scalar field. 
In this case, the Reissner-Nordstr\"{o}m black holes can 
evolve into perturbatively stable hairy solutions. 
Moreover, the occurrence of spontaneous scalarization is
not necessarily restricted to strong gravity backgrounds. 
Even in the absence of gravity, the coupling 
$-\alpha(\phi) F_{\mu \nu}F^{\mu \nu}/4$ 
is capable of inducing scalarization of a charged object like 
a conducting sphere \cite{Herdeiro1,Herdeiro:2020htm}.
Essentially, the notion of spontaneous scalarization
is not exclusive for black holes/neutron stars
and for the presence of gravity in underlying theories,
and the external strong gravitational or electric field
just acts as the trigger of tachyonic instabilities 
in the scalar-field sector.

The past works about charge-induced spontaneous scalarization 
have been restricted to black holes or conducting charged spheres 
in Minkowski spacetime. 
In this paper, we will study whether spontaneous scalarization 
can take place for electrically charged stars in Einstein-Maxwell-scalar 
theories. 
For the matter sector inside the star, we consider a perfect fluid with a 
matter density $\rho_m$ endowed with a conserved charged current 
${J_{c}}^{\mu}$ coupled to the 
gauge field $A_{\mu}$. In GR, it is known that the extra 
Coulomb force induced by a large charge density $\rho_c$ 
leads to an imbalance between gravity and 
fluid pressures \cite{Ray:2003gt} (see also 
Refs.~\cite{Bekenstein:1971ej,deFelice:1999qp,Anninos:2001yb}). 
Introducing the ratio $\mu:=M_{\rm pl}\left(\rho_c/\rho_m\right)$, 
charged stars in GR tend to be unstable as $\mu$ 
approaches its maximally allowed value around 0.7.

Our goal in this paper is to elucidate whether the scalar-gauge coupling 
$-\alpha(\phi) F_{\mu \nu}F^{\mu \nu}/4$ gives rise to scalarized 
solutions which can be the endpoint of tachyonic instabilities of 
the GR branch $\phi=0$.
For the coupling $\alpha (\phi)=\exp (-\beta \phi^2/M_{\rm pl}^2)$, 
we will show that, above some threshold values of $\mu$,  
the scalarized stars whose masses and radii are 
smaller than those in the GR branch arise for $\beta=-{\cal O}(1)$ 
in the wide range of matter densities. 
Thus, charge-induced spontaneous scalarization can 
occur not only for black holes but also for gravitationally 
bounded stars.

Throughout the paper, we will work in natural units with
the reduced Planck mass $M_{\rm pl}$.
If necessary, one can switch to Gaussian-cgs units 
with the gravitational constant $G$
and the speed of light $c$,
by the replacements of 
 $M_{\rm pl}\to c^2/\sqrt{8\pi G}$,
$\rho_m\to \rho_m c^2$,
and $\rho_c\to\sqrt{8\pi}\rho_c$,
with $\mu \to (1/\sqrt{G})\left(\rho_c/\rho_m\right)$.

%%%%%%%%%%%%%%%%%%%%%%%%%%%%%%%%%%%%%%%%%%
\section{Einstein-Maxwell-scalar theories with a charged perfect fluid}
\label{basicsec}
%%%%%%%%%%%%%%%%%%%%%%%%%%%%%%%%%%%%%%%%%%

We begin with Einstein-Maxwell-scalar theories given by the action 
\be
{\cal S} =
\int {\rm d}^4 x 
\left[ \frac{M_{\rm pl}^2}{2}\sqrt{-g}R
+L_{\phi F}
+L_{m}
+L_c
\right]\,,
\label{action}
\ee
where
\begin{eqnarray}
\label{lpf}
L_{\phi F}
&:=&
-\sqrt{-g}
\left[
\frac{1}{2} \nabla_{\mu} \phi  \nabla^{\mu} \phi
+\frac{\alpha(\phi)}{4} F_{\mu \nu} F^{\mu \nu}
\right]\,,
\\
\label{lm}
L_m
&:=& -\left[ \sqrt{-g}\,\rho_m (n) +{J_m}^{\mu}\partial_{\mu}\ell_m \right]\,,
\\
\label{lc}
L_c
&:=&
-{J_c}^{\mu} 
\left( \partial_{\mu}\ell_c-A_{\mu} \right)
=
-J_{c \mu} \left( \partial_{\nu}\ell_c-A_{\nu} \right)g^{\mu \nu}\,.
\end{eqnarray}
Here, $g$ is the determinant of metric tensor $g_{\mu \nu}$, 
$R$ is the Ricci scalar, $\alpha(\phi)$ is a function of 
the scalar field $\phi$, and
$F_{\mu \nu}:=\nabla_{\mu}A_{\nu}-\nabla_{\nu}A_{\mu}
\,(=\partial_\mu A_\nu-\partial_\nu A_\mu)$ 
is the antisymmetric field strength tensor of a gauge field $A_{\mu}$, 
with $\nabla_{\mu}$ the covariant derivative operator,
respectively. 

The term $L_m$ given by Eq.~\eqref{lm} corresponds to the 
Schutz-Sorkin Lagrangian \cite{Sorkin,Brown,DGS,Kase:2020hst} 
describing a perfect fluid with the matter 
density $\rho_m$ and current vector ${J_m}^{\mu}$. 
The matter density $\rho_m$ is a function of 
the fluid number density $n$.
The scalar quantity $\ell_m$ is a Lagrange multiplier, 
where $\partial_{\mu}\ell_m$ is the partial derivative of 
$\ell_m$ with respect to the spacetime coordinate $x^{\mu}$.
The existence of the term $\partial_{\mu}\ell_m$ 
in Eq.~(\ref{action}) ensures the current conservation
\be
\partial_{\mu} {J_m}^{\mu}=0\,.
\label{cuJm}
\ee
Besides the derivative $\partial_{\mu} \ell_m$, we generally have the term
${\cal A}_i \partial_{\mu}{\cal B}^i$ arising from spatial vectors 
${\cal A}_i$ and ${\cal B}^i$. 
Since ${\cal A}_i$ vanishes on a static and spherically symmetric 
background \cite{Kase:2020qvz}, we will not consider this term throughout the paper. 
The fluid four velocity $u^{\mu}$ is related to ${J_m}^{\mu}$, as
\be
{J_m}^{\mu}=n\sqrt{-g}\,
u^{\mu}\,.
\label{Jm1}
\ee
Due to the property $u^{\mu}u_{\mu}=-1$, the number density 
can be expressed in the form 
\be
n=\sqrt{\frac{{J_m}^{\mu}{J_{m\mu}}}{g}}\,.
\label{Jm2}
\ee
In the Lagrangian $L_c$ given by Eq.~\eqref{lc}, 
${J_c}^{\mu}$ corresponds to a charged current.
The presence of a Lagrange multiplier $\partial_{\mu} \ell_c$ ensures 
the charge current conservation 
\be
\partial_{\mu} {J_c}^{\mu}=0\,.
\ee
Analogous to Eqs.~(\ref{Jm1}) and (\ref{Jm2}), 
there are the following relations 
\be
{J_c}^{\mu}=\rho_c \sqrt{-g}\,u^{\mu}\,,\qquad 
\rho_c=\sqrt{\frac{{J_c}^{\mu}{J_{c\mu}}}{g}}\,,
\ee
where $\rho_c$ is a charge density.
Varying the action (\ref{action}) with respect to 
${J_c}^{\mu}$, 
it follows that 
\be
\partial_{\mu}\ell_c=A_{\mu}\,.
\label{ellc}
\ee
Variations of the action (\ref{action}) with respect to $A_\mu$ 
and $\phi$ lead, respcetively, to
\ba
& &
\nabla_{\mu} \left[ \alpha(\phi) F^{\mu \nu} \right]
=-\frac{{J_c}^{\nu}}{\sqrt{-g}}
=-\rho_c u^{\nu}\,,
\label{Aeq1}\\
& & g_{\mu \nu} \nabla^{\mu}\nabla^{\nu}\phi
-\frac{\alpha_{,\phi}}{4} 
F_{\mu \nu} F^{\mu \nu}=0\,,
\label{phieq1}
\ea
where $\alpha_{,\phi}:={\rm d} \alpha/{\rm d}\phi$.

Varying the Lagrangian $L_{\phi F}$ given by Eq.~\eqref{lpf}
with respect to $g^{\mu \nu}$ and using the 
property $\delta \sqrt{-g} = -\sqrt{-g}
g_{\mu \nu} \delta g^{\mu \nu}/2$, the energy-momentum 
tensor arising from $L_{\phi F}$ yields
\be
T_{\mu \nu}^{(\phi F)}=
-\frac{2}{\sqrt{-g}} \frac{\delta L_{\phi F}}{\delta g^{\mu \nu}}=
\nabla_{\mu}\phi \nabla_{\nu}\phi
-\frac{1}{2} g_{\mu \nu} \nabla_{\alpha}\phi \nabla^{\alpha}\phi
+\alpha(\phi) g^{\rho \sigma} 
F_{\mu \rho} F_{\nu \sigma}-\frac{\alpha(\phi)}{4} g_{\mu \nu} 
F^{\rho \sigma} F_{\rho \sigma}\,.
\ee
We also vary the Schtuz-Sorkin Lagrangian $L_m$ given by Eq.~\eqref{lm}
with respect to $g^{\mu \nu}$ and exploit the property
$\delta n = (n/2)\left( g_{\mu \nu} 
-u_{\mu} u_{\nu} \right) \delta g^{\mu \nu}$. 
We also make use of the relation  
$\partial_{\mu}\ell_m=\rho_{m,n}u_{\mu}$, which follows from
the variation of Eq.~(\ref{action}) with respect to 
${J_m}^{\mu}$ \cite{Kase:2020qvz,Tsujikawa:2020die}.
Then, we obtain the standard form of the 
perfect-fluid energy-momentum tensor  
\be
T_{\mu \nu}^{(m)}=
-\frac{2}{\sqrt{-g}} \frac{\delta L_m}{\delta g^{\mu \nu}}
=\left( \rho_m+P_m \right) u_{\mu} u_{\nu}+P_m g_{\mu \nu}\,,
\label{Tmunu}
\ee
where $P_m$ is the matter pressure defined by 
\be
P_m=n \rho_{m,n}-\rho_m\,.
\ee

For the last term of Eq.~(\ref{action}), we vary  
the Lagrangian $L_c$ given by Eq.~\eqref{lc} with respect to $g^{\mu \nu}$. 
On using Eq.~(\ref{ellc}), the energy-momentum tensor associated 
with $L_c$ is
\be
T_{\mu \nu}^{(c)}=
-\frac{2}{\sqrt{-g}} \frac{\delta L_c}{\delta g^{\mu \nu}}
=0\,.
\label{Tmunuc}
\ee

Then, the gravitational field equations arising from the action 
(\ref{action}) with Eqs.~\eqref{lpf}-\eqref{lc} are
\be
M_{\rm pl}^2 G_{\mu \nu} 
=T_{\mu \nu}^{(\phi F)}+T_{\mu \nu}^{(m)}\,,
\label{graeq}
\ee
where $G_{\mu \nu}$ is the Einstein tensor obeying the 
Bianchi identity $\nabla^{\nu} G_{\mu\nu}=0$.
On using Eqs.~(\ref{Aeq1}) and (\ref{phieq1}) as well as 
the Maxwell equations 
$\nabla_{\mu}F_{\nu \sigma}+\nabla_{\nu}F_{\sigma \mu}
+\nabla_{\sigma}F_{\mu \nu}=0$, it follows that  
\be
\nabla^{\nu}T_{\mu \nu}^{(\phi F)}
=-\rho_c F_{\mu \nu} u^{\nu}\,.
\ee
Due to the conservation of the total energy-momentum tensor 
$\nabla^{\nu}T_{\mu \nu}^{(\phi F)}+\nabla^{\nu}T_{\mu \nu}^{(m)}=0$, 
we have
\be
\nabla^{\nu}T_{\mu \nu}^{(m)}
=\rho_c F_{\mu \nu} u^{\nu}\,,
\label{Tmunum}
\ee
where the right hand side corresponds to the Coulomb force. 
Since the anti-symmetric tensor $F_{\mu \nu}$ satisfies the relation 
$F_{\mu \nu} u^{\mu}u^{\nu}=0$, multiplying Eq.~(\ref{Tmunum})
with $u^{\mu}$ gives 
\be
u^{\mu}\nabla^{\nu}T_{\mu \nu}^{(m)}=0\,.
\label{Tmunum2}
\ee
{}From Eq.~(\ref{Jm1}), the matter current 
conservation (\ref{cuJm}) translates to 
\be
u_{\mu}\nabla^{\mu}\rho_m
+(\rho_m+P_m)\nabla^{\mu}u_{\mu}=0\,.
\label{cucon2}
\ee
On using (\ref{cucon2}), Eq.~(\ref{Tmunum}) can be expressed 
in the form 
\be
\left( g_{\mu \nu}+u_{\mu} u_{\nu} \right)
\nabla^\nu P_m
+\left(\rho_m+P_m\right)
u_\nu \nabla^\nu u_\mu
=\rho_c F_{\mu \nu} u^{\nu}\,.
\label{Tmunum3}
\ee
Multiplying the left hand side of Eq.~(\ref{Tmunum3}) with 
$u^{\mu}$ and exploiting the property $u^{\mu} u_{\mu}=-1$, 
we can confirm that the relation (\ref{Tmunum2}) indeed holds.
We also introduce a unit vector $n^{\nu}$ orthogonal to 
$u_{\nu}$, such that  
$n^{\nu}u_{\nu}=0$ and $n^{\nu}n_{\nu}=1$. 
Multiplying Eq.~(\ref{Tmunum3}) with $n^{\mu}$, we obtain
\be
n_{\nu} \nabla^{\nu} P_m
+(\rho_m+P_m)n^{\mu}u_{\nu} \nabla^{\nu} u_{\mu}=
\rho_c F_{\mu\nu}u^\nu n^\mu\,.
\label{nPm}
\ee
This shows the balance between the three forces, i.e., 
the pressure gradient, gravity, and Coulomb force along the 
$n^{\mu}$ direction.

%%%%%%%%%%%%%%%%%%%%%%%%%%%%%%%%%%%%%%%%%%
\section{Static and spherically symmetric background}
\label{backsec}
%%%%%%%%%%%%%%%%%%%%%%%%%%%%%%%%%%%%%%%%%%

Let us consider a static and spherically symmetric background 
given by the line element
\be
{\rm d}s^2=-f(r){\rm d}t^2+h^{-1} (r){\rm d}r^2+r^2
\left( {\rm d}\theta^2+\sin^2 \theta\,{\rm d}\varphi^2 \right)\,,
\label{metric}
\ee
where $f(r)$ and $h(r)$ are functions of the radial 
coordinate $r$. 
The configurations of $\phi$, $n$, $A_{\mu}$, and $u^{\mu}$ 
compatible with this background are given by 
\be
\phi=\phi(r)\,,\qquad 
n=n(r)\,,\qquad
A_{\mu}=\left( A_t (r), 0, 0, 0 \right)\,, \qquad
u^{\mu}=\left( f(r)^{-1/2}, 0, 0, 0 \right)\,.
\ee

Then, Eqs.~(\ref{Aeq1}) and (\ref{phieq1}) reduce, 
respectively, to  
\ba
& &
A_t''+\left( \frac{2}{r}-\frac{f'}{2f}+\frac{h'}{2h}
+\frac{\alpha_{,\phi}}{\alpha} \phi' \right) A_t'
-\frac{\sqrt{f}\rho_c}{h \alpha}=0\,,
\label{Ateq}\\
& &
\phi''+\left( \frac{2}{r}+\frac{f'}{2f}+\frac{h'}{2h}
\right) \phi'+\frac{A_t'^2}{2f}\alpha_{,\phi}=0\,,
\label{phieq}
\ea
where a `prime' represents the derivative with respect to $r$.
The $tt$ and $rr$ components of Einstein Eqs.~(\ref{graeq}) give
\ba
h' &=& -\frac{h-1}{r}-\frac{h r\phi'^2}{2M_{\rm pl}^2}
-\frac{\alpha r h A_t'^2}{2f M_{\rm pl}^2}
-\frac{r}{M_{\rm pl}^2} \rho_m \,,\label{heq}\\
f' &=& -\frac{f(h-1)}{hr}+\frac{f r\phi'^2}{2M_{\rm pl}^2}
-\frac{\alpha r A_t'^2}{2 M_{\rm pl}^2}
+\frac{fr}{h M_{\rm pl}^2}P_m\,.
\label{fdeq}
\ea
Taking the unit vector orthogonal to $u_{\mu}$ as 
$n^{\mu}=(0,h^{1/2}(r),0,0)$, Eq.~(\ref{nPm}) reduces to
\be
P_m'+\frac{f'}{2f} \left( \rho_m+P_m \right)
=\frac{\rho_c A_t'}{\sqrt{f}}\,.
\label{Peq}
\ee
The star radius $r_s$ is identified by the condition  
\be
P_m (r_s)=0\,.
\ee
We assume that $\rho_m$, $P_m$, and $\rho_c$ vanish 
outside the star. Inside the star, we define the 
dimensionless ratio
\be
\label{def_mu}
\mu:=M_{\rm pl}
\frac{ \rho_c}{\rho_m}\,,
\ee
which generally depends on $r$. 
For the numerical analysis performed later in Sec.~\ref{sponsec}, 
we will consider the case in which $\mu$ is constant.

We define the mass function ${\cal M}(r)$, according to 
\be
h(r)=1-\frac{{\cal M}(r)}{4\pi M_{\rm pl}^2 r}\,.
\ee
The Arnowitt-Deser-Misner (ADM) mass corresponds to  
the asymptotic value of ${\cal M}(r)$, such that  
\be
M=\lim_{r \to \infty} {\cal M}(r)=\lim_{r \to \infty}4\pi M_{\rm pl}^2 r \left[ 1-h(r) 
\right]\,.
\ee
As we see in Eq.~(\ref{heq}), the ADM mass not only contains the contribution 
from $\rho_m$ but also those from $A_t'$ and $\phi'$. 
Since both $A_t'$ and $\phi'$ do not vanish outside the star, the mass 
${\cal M}(r_s)$ computed at the surface of star is not generally 
identical to $M$. 
To extract the mass from the matter density $\rho_m$, we define the 
proper mass of star, as 
\be
M_p := \int_0^{\infty} {\rm d}^3x\, \rho_m 
\sqrt{{}^{(3)}g}
=\int_0^{r_s} \frac{4\pi \rho_m\,r^2}{\sqrt{h}}{\rm d}r\,,
\label{Mp}
\ee
where ${}^{(3)}g$ is the determinant of 
three-dimensional spatial metric. 
We define the gravitational binding energy, according to 
the difference between $M_p$ and $M$, as
\be
\Delta := M_p-M\,.
\label{Delta}
\ee
The star is gravitationally bound for $\Delta>0$, 
which can be regarded as a necessary condition for 
its perturbative stability,
while if $\Delta<0$ the star is gravitationally unbounded. 

For the equation of state (EOS) of relativistic stars, 
we consider a polytropic type of 
the form \cite{Damour:1993hw}
\be
\rho_m=\bar{\rho}_m \left( \chi+\frac{K}{\Gamma-1}
\chi^{\Gamma} \right)\,,\qquad 
P_m=K\bar{\rho}_m \chi^{\Gamma}\,,
\label{eos}
\ee
where $\bar{\rho}_m$, $K$, and $\Gamma$ are constants. 
We define the constant matter density $\bar{\rho}_m$ as 
$\bar{\rho}_m=\bar{n}_m m_b=1.6749 \times 10^{14}$~g~cm$^{-3}$, 
where $\bar{n}_m=0.1\,{\rm (fm)^{-3}}$ is 
the typical nuclear number density and $m_b$ is
the mean rest mass of baryons.
The dimensionless variable $\chi$ is given by $\chi=n_b/\bar{n}_m$, 
where $n_b$ is the baryon number density. 
The EOS parameter is defined by 
\be
w_m:=\frac{P_m}{\rho_m}
=\frac{K \chi^{\Gamma-1}}
{1+K\chi^{\Gamma-1}/(\Gamma-1)}\,.
\label{wpoly}
\ee
In the low-density regime characterized by $K \chi^{\Gamma-1} \ll 1$ 
the EOS parameter reduces to $w_m \simeq K \chi^{\Gamma-1}$, whereas, 
in the high-density regime ($K \chi^{\Gamma-1} \gg 1$), $w_m \simeq \Gamma-1$.

To solve the background Eqs.~(\ref{Ateq})-(\ref{Peq}) numerically, 
we introduce the following dimensionless quantities
\be
s= \ln \frac{r}{\bar{r}}\,,\qquad 
y=\frac{\rho_m}{\bar{\rho}_m}\,,\qquad 
z=\frac{P_m}{\bar{\rho}_m}\,,\qquad 
\tilde{A}_t=\frac{A_t}{M_{\rm pl}}\,,\qquad 
\tilde{\phi}=\frac{\phi}{M_{\rm pl}}\,,\qquad
m(r)=\frac{3{\cal M}(r)}{4\pi \bar{\rho}_m \bar{r}^3}\,,
\label{wdef}
\ee
where 
\be
\bar{r}=\frac{M_{\rm pl}}{\sqrt{\bar{\rho}_m}}
=17.885~{\rm km}\,.
\ee
Then, Eq.~(\ref{Peq}) can be expressed in the form 
\be
\frac{{\rm d}z}{{\rm d}s}+\frac{1}{2f} \frac{{\rm d}f}{{\rm d}s}
\left( y+z \right)=\frac{\mu y}{\sqrt{f}} 
\frac{{\rm d} \tilde{A}_t}{{\rm d} s}\,,
\label{zeq}
\ee
where, for the polytropic EOS, we have 
\be
y=\left( \frac{z}{K} \right)^{1/\Gamma}+\frac{z}{\Gamma-1}\,,
\qquad 
z=K \chi^{\Gamma}\,.
\label{yz}
\ee
In terms of the solar mass $M_{\odot}=1.9884 \times 10^{33}$\,g, 
the mass function ${\cal M}(r)$ is expressed as 
\be
{\cal M}(r)=2.0186\,m (r) M_{\odot}\,,
\label{Mrre}
\ee
where, from Eq.~(\ref{heq}), $m(r)$ obeys
\be
\frac{{\rm d}m}{{\rm d}s}=3y e^{3s}+\frac{3x-m}{2f} 
\left[ f \left( \frac{{\rm d} \tilde{\phi}}{{\rm d}s} 
\right)^2+\alpha \left( \frac{{\rm d} \tilde{A}_t}{{\rm d}s} 
\right)^2 \right]\,.
\label{mdeq}
\ee
For a given value of $\chi$ around $r=0$,
we integrate Eqs.~(\ref{zeq}) and (\ref{mdeq}) outwards
together with the normalized versions of 
Eqs.~(\ref{Ateq}), (\ref{phieq}), and 
(\ref{fdeq}) to solve for $z$, $m$, $\tilde{A}_t$, $\tilde{\phi}$, 
and $f$. The ADM mass $M$ is known from Eq.~(\ref{Mrre}) 
after the integration to a distance much larger than $r_s$.
We also define ${\cal M}_p(r)=4\pi \bar{\rho}_m \bar{r}^3 m_{p}(r)/3=
\int_0^r 4\pi \rho_m \tilde{r}^2/\sqrt{h}\,{\rm d}\tilde{r}$ 
and solve the differential equation 
\be
\frac{{\rm d} m_p}{{\rm d}s}=\frac{3y e^{3s}}{\sqrt{h}}\,,
\ee
up to $r=r_s$. Then, we obtain the proper mass of star (\ref{Mp}),
according to the relation $M_p=2.0186 m_p(r_s) M_{\odot}$.

%%%%%%%%%%%%%%%%%%%%%%%%%%%%%%%%%%%%%%%%%%
\section{Spontaneous scalarization of charged stars}
\label{sponsec}
%%%%%%%%%%%%%%%%%%%%%%%%%%%%%%%%%%%%%%%%%%

In this section we first derive a necessary condition for spontaneous scalarization 
to occur and then study the existence of scalarized solutions for a concrete 
scalar-gauge coupling. Varying the action (\ref{action}) 
with respect to $\phi$ on a general background, 
the scalar field obeys
\be
\square \phi+\alpha_{,\phi}(\phi) F=0\,,\qquad {\rm where} 
\qquad F=-\frac{1}{4} F_{\mu \nu} F^{\mu \nu}\,.
\ee
First, we require that, in the absence of the scalar field, $\phi=0$, 
the standard normalization for the Maxwell term
$- F_{\mu \nu} F^{\mu \nu}/4$
is recovered in the gravitational action \eqref{action}, i.e., $\alpha(0)=1$.
The existence of the GR branch $\phi=0$ requires 
that $\alpha_{,\phi} (0)=0$. 
In this paper, by ``GR solutions''
we mean the solutions in Einstein-Maxwell theory
minimally coupled to the matter sector.

The perturbation $\delta \phi$ around the solution $\phi=0$ obeys
\be
\left( \square -m_{\rm eff}^2 \right) \delta \phi=0\,,\qquad 
{\rm where} \qquad m_{\rm eff}^2=-F \alpha_{,\phi \phi}(0)\,.
\ee
There is a tachyonic instability for $m_{\rm eff}^2<0$, 
i.e., $F\alpha_{,\phi \phi}(0)>0$. 
On the background (\ref{metric}),
we have $F=h A_t'^2/(2f)>0$, 
so the solution $\phi=0$ is unstable for $\alpha_{,\phi \phi}(0)>0$.
In summary, the necessary conditions for spontaneous scalarization 
to occur from the GR branch to the other nontrivial 
branch with $\phi \neq 0$ are given by 
\be
\alpha_{,\phi} (0)=0\,,\qquad 
\alpha_{,\phi \phi}(0)>0\,.
\label{necon}
\ee
To satisfy these conditions, the coupling $\alpha (\phi)$ needs to 
take the form
\begin{eqnarray}
\label{generic_coupling}
\alpha (\phi)=1 -\beta \frac{\phi^2}{M_{\rm pl}^2}
+\sum_{n=2}^\infty \beta_n \left(\frac{\phi^2}{M_{\rm pl}^2}\right)^{n}, 
\end{eqnarray}
where $\beta$ and $\beta_n$ ($n=2, 3, 4, \cdots$) are constants. 
One of the examples is 
\be
\alpha (\phi)=\exp \left(-\beta \frac{\phi^2}{M_{\rm pl}^2}\right)\,.
\label{alpha}
\ee
In this case, the above necessary 
conditions are satisfied for 
\be
\beta<0\,.
\ee
In the following, we will focus on the coupling (\ref{alpha}) 
to study the existence of scalarized solutions. 

\subsection{
Expansions in the vicinity of center $r=0$ and 
spatial infinity $r \to \infty$}

Around the origin of star ($r=0$), we need to impose the regular 
boundary conditions 
$\phi'(0)=A_t'(0)=h'(0)=f'(0)=\rho_m'(0)=P_m'(0)=\rho_c'(0)=0$. 
The solutions compatible with these conditions 
are expressed in the forms of
\ba
& &
\phi(r)=\phi_0+\sum_{i=2}^{\infty} \phi_i r^i\,,\qquad 
A_t(r)=A_{t0}+\sum_{i=2}^{\infty} A_{ti} r^i\,,\qquad 
h(r)=1+\sum_{i=2}^{\infty} h_i r^i\,,\qquad 
f(r)=f_0+\sum_{i=2}^{\infty} f_i r^i\,,\nonumber \\
& &
\rho_m (r)=\rho_{m0}+\sum_{i=2}^{\infty} \rho_{mi} r^i\,,\qquad 
P_m (r)=P_{m0}+\sum_{i=2}^{\infty} P_{mi} r^i\,,\qquad 
\rho_c (r)=\rho_{c0}+\sum_{i=2}^{\infty} \rho_{ci} r^i\,,
\label{r=0exp}
\ea
where $\phi_0, \phi_i, A_{t0}, A_{ti}, h_i, f_0, f_i, 
\rho_{m0}, \rho_{mi}, P_{m0}, P_{mi}, 
\rho_{c0}$, and $\rho_{ci}$ are constants.
Substituting Eq.~(\ref{r=0exp}) into Eqs.~(\ref{Ateq})-(\ref{Peq}), 
the iterative solutions, up to next-to-leading order, are given by 
\ba
\phi(r) &=&\phi_0+\frac{\beta \phi_0 e^{\beta \phi_0^2/M_{\rm pl}^2} \rho_{c0}^2}
{180M_{\rm pl}^2}r^4+{\cal O} (r^6)\,,
\label{phir=0}\\
A_t (r) &=& A_{t0}+\frac{\sqrt{f_0} \rho_{c0} e^{\beta \phi_0^2/M_{\rm pl}^2}}{6}r^2
+{\cal O} (r^4)\,,
\label{Atr=0}\\
h(r) &=& 1-\frac{\rho_{m0}}{3M_{\rm pl}^2}r^2
+{\cal O} (r^4)\,,\\
f(r) &=& f_0+\frac{f_0(\rho_{m0}+3P_{m0})}
{6M_{\rm pl}^2}r^2+{\cal O} (r^4)\,,\\
P_m(r) &=& P_{m0}-\frac{(\rho_{m0}+3P_{m0})
(\rho_{m0}+P_{m0})-2 e^{\beta \phi_0^2/M_{\rm pl}^2}
\rho_{c0}^2M_{\rm pl}^2}
{12M_{\rm pl}^2}r^2+{\cal O} (r^4)\,.
\label{Pm}
\ea
The constant $f_0$ represents the freedom of 
rescaling of the time coordinate,
whose choice does not affect physical
quantities such as the mass and radius of star.

When $\beta=0$ or $\phi_0=0$,
the scalar field $\phi(r)$ vanishes, 
so that it does not affect the profiles of $A_t(r)$, $h(r)$, $f(r)$, 
and $P_m (r)$. 
The existence of charge density $\rho_{c0}$ gives rise to 
a nonvanishing electric field $A_t'(r)=\sqrt{f_0} \rho_{c0}r/3+{\cal O}(r^3)$ 
around the origin.
For $\beta=0$ or $\phi_0=0$, the last term 
$\rho_{c0}^2 r^2/6$ in Eq.~(\ref{Pm}) 
works as a Coulomb force against gravity. 
In the absence of this term, the pressure decreases for increasing $r$. 
In order to ensure the existence of charged stars 
against the destabilization by the Coulomb force,  
we require that $P_m'(r)<0$. 
We use the relation $\rho_{c0}=\rho_{m0} \mu/M_{\rm pl}$
obtained from Eq.~\eqref{def_mu} and assume that 
$\mu$ is constant. Then, from Eq.~(\ref{Pm}), the condition 
$P_m'(r)<0$ translates to 
\be
\mu<\sqrt{\frac{(1+3w_{m0})(1+w_{m0})}{2}}\,,
\label{muup}
\ee
where $w_{m0}=P_{m0}/\rho_{m0}$ is the EOS parameter at $r=0$.
To satisfy this condition in the nonrelativistic regime with $w_{m0} \ll 1$, 
the parameter $\mu$ should be in the range
\be
\mu<\frac{1}{\sqrt{2}}\,.
\label{muup2}
\ee
Thus, the charge density relative to the matter density is 
bounded from above to realize a gravitationally 
bounded star.

When $\beta<0$ and $\phi_0\neq 0$, 
the second term on the right hand side of 
Eq.~(\ref{phir=0}), which is proportional to $r^4$,  
leads to the variation of $\phi(r)$. 
Around $r=0$, the metric components $h(r)$ and $f(r)$ 
do not possess the $\phi_0$ dependence up to the order of $r^2$, 
but the scalar field affects $P_m (r)$ through the 
Coulomb force $e^{\beta \phi_0^2/M_{\rm pl}^2} \rho_{c0}^2r^2/6$. 
For the negative $\beta$ we have $e^{\beta \phi_0^2/M_{\rm pl}^2}<1$, 
so the Coulomb force is suppressed in comparison to 
the case of $\beta=0$.
This suggests that the scalar-gauge coupling with $\beta<0$ 
may stabilize the charged star even for the values of $\mu$ 
close to the upper bound (\ref{muup}).
The iterative solutions (\ref{phir=0})-(\ref{Pm}) lose 
their validity around the surface of star, so we will numerically 
integrate Eqs.~(\ref{Ateq})-(\ref{Peq}) around from $r=0$ up to 
a sufficiently large $r$ to confirm the 
existence of hairy solutions with $\phi(r)\neq 0$.

Outside the star ($r>r_s$), we have $\rho_c=0$, $\rho_m=0$, 
and $P_m=0$ in Eqs.~(\ref{Ateq})-(\ref{Peq}). 
Then, the integrated solution to Eq.~(\ref{Ateq}) is 
expressed in the form
\be
A_t'(r)=Q \frac{\sqrt{f(r)}\,
e^{\beta \phi^2 (r)/M_{\rm pl}^2}}
{r^2 \sqrt{h(r)}}\,,
\label{Atso}
\ee
where $Q$ is a constant. 
We will consider the case in which the boundary conditions 
at spatial infinity are given by 
\be
\phi (\infty)=0\,,\qquad 
h(\infty)=1\,,\qquad f(\infty)=1\,.
\label{bofh}
\ee
Then, for $r \gg r_s$, Eq.~(\ref{Atso}) has the radial dependence 
$A_t'(r) \simeq Q/r^2$. 
Neglecting the term $A_t'^2 \alpha_{,\phi}/(2f)$ in Eq.~(\ref{phieq}) 
and integrating this equation, we obtain
\be
\phi'(r)=\frac{c_1}{r^2 \sqrt{f(r)h(r)}}\,,
\label{phiso}
\ee
where $c_1$ is a constant.
At the distance $r \gg r_s$ the term $A_t'^2 \alpha_{,\phi}/(2f)$ 
decreases faster than $1/r^4$, so we can ignore its contribution 
to Eq.~(\ref{phieq}). 
Substituting the solutions (\ref{Atso}) and (\ref{phiso}) into
Eqs.~(\ref{heq}) and (\ref{fdeq}), it follows that 
the contributions of $A_t'(r)$ and $\phi'(r)$ to $f(r)$ and $h(r)$ 
work as corrections (proportional to $1/r^2$) to the leading-order 
solutions $f(r)=h(r)=1-M/(4\pi M_{\rm pl}^2 r)$. 

The two asymptotic solutions derived in the regimes $r \ll r_s$ and 
$r \gg r_s$ should be matched around the surface of star. 
Since we are considering $U(1)$ gauge-invariant 
theory, the contributions arising from the vector field 
to Eqs.~(\ref{Ateq})-(\ref{Peq}) are the derivatives of $A_t$ alone.
Hence, 
we do not need to specify the constant $A_{t0}$ in Eq.~(\ref{Atr=0}). 
On the other hand, the field value $\phi$ at $r=0$, i.e., $\phi_0$,
should be determined to satisfy the boundary condition $\phi (\infty)=0$ 
at spatial infinity. 
We will iteratively identify the value of $\phi_0$ by solving the 
background equations of motion numerically.

\subsection{No-hair solutions with $\phi=0$}
\label{nohairsec}

Let us first revisit compact star solutions 
with the trivial scalar field $\phi=0$. 
In this case the term $A_t'^2 \alpha_{,\phi}/(2f)$ vanishes even inside 
the star for the coupling \eqref{alpha}, so Eq.~(\ref{phieq}) 
is trivially satisfied for any value of $\beta$. 
{}From Eq.~(\ref{Atso}), the solution to $A_t'(r)$ outside the star 
($r>r_s$) is given by  
\be
A_t'(r)=Q \frac{\sqrt{f(r)}}{r^2 \sqrt{h(r)}}\,.
\label{Atr}
\ee
Substituting Eq.~(\ref{Atr}), $\phi(r)=\phi'(r)=0$, 
and $\rho_m=0=P_m$ 
into Eqs.~(\ref{heq}) and (\ref{fdeq}), 
the metric components outside the star consistent with 
boundary conditions (\ref{bofh}) are 
\be
f=h=1-\frac{M}{4\pi M_{\rm pl}^2 r}
+\frac{Q^2}{2M_{\rm pl}^2 r^2}\,,
\ee
where $M$ corresponds to the ADM mass.
This is known as the Reissner-Nordstr\"{o}m metric
in the Einstein-Maxwell model.
Thus, the GR solution exists for any value of $\beta$.

\begin{figure}[h]
\begin{center}
\includegraphics[height=3.25in,width=3.4in]{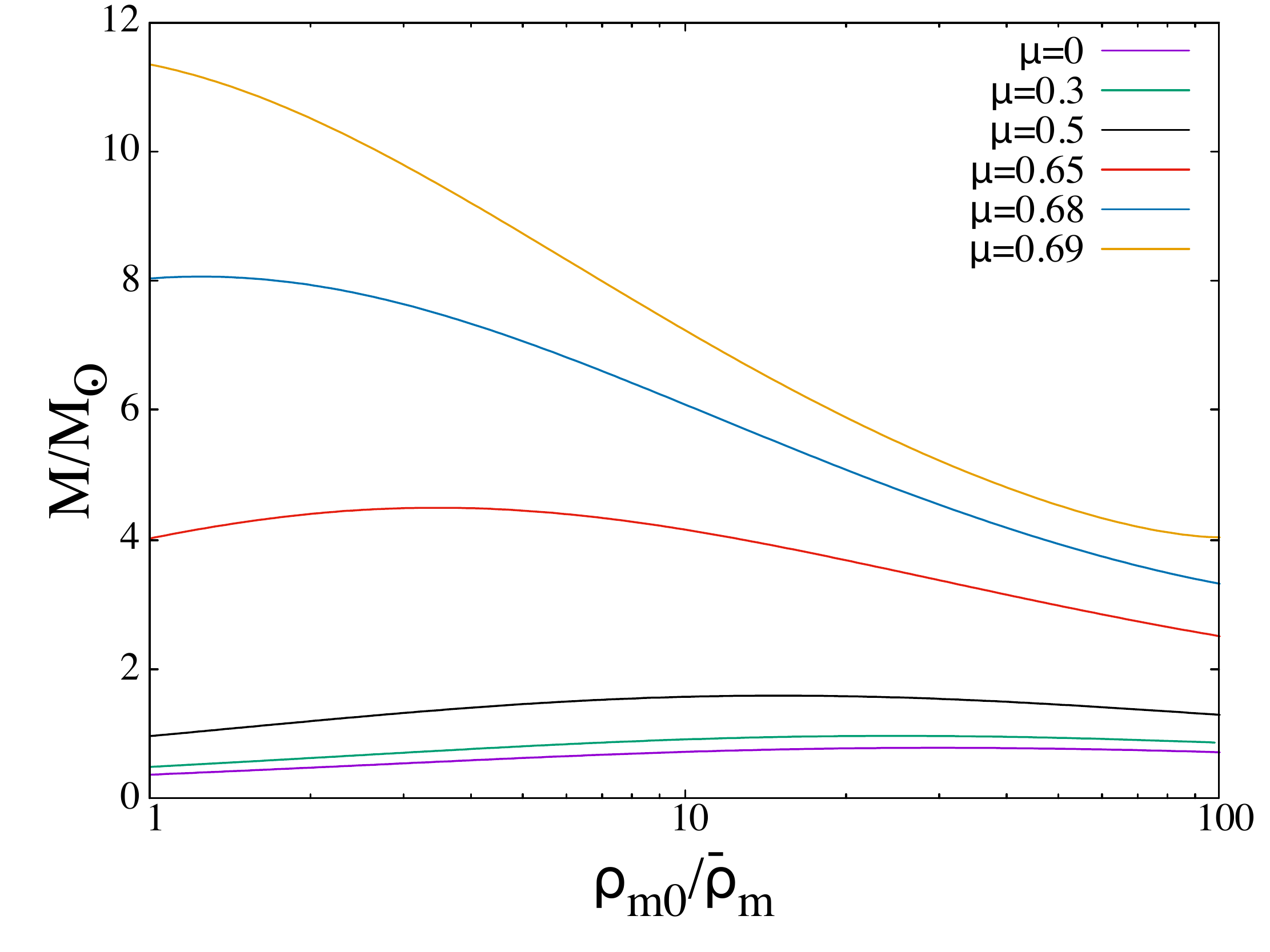}
\includegraphics[height=3.62in,width=3.4in]{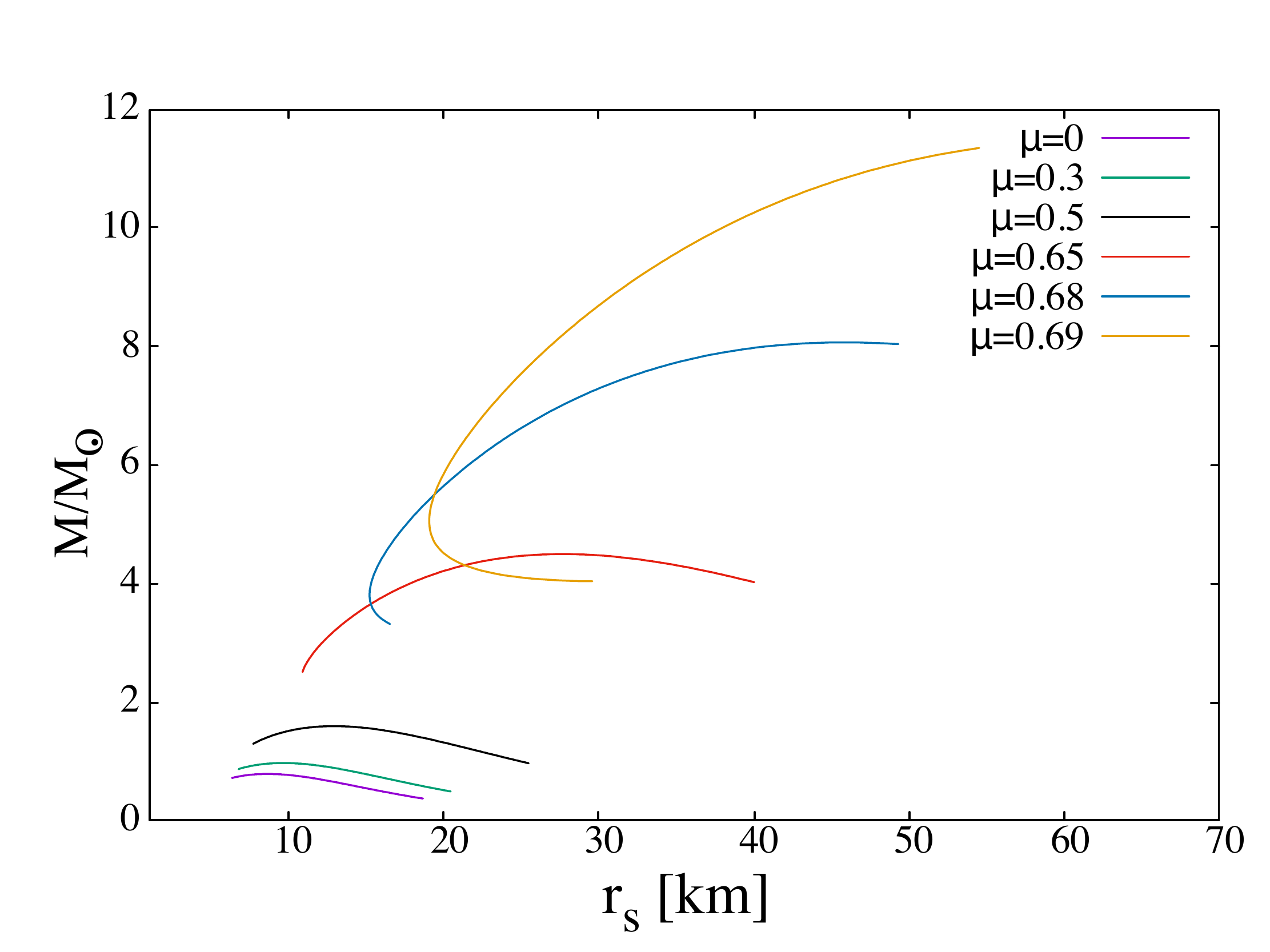}
\end{center}
\caption{(Left) The ADM mass $M$ (normalized by the solar mass $M_{\odot}$) 
versus the central mass density $\rho_{m0}$ (normalized by 
$\bar{\rho}_m=1.6749 \times 10^{14}$ g~cm$^{-3}$) 
for $\beta=0$. We choose the polytropic EOS with 
$\Gamma=5/3$ and $K=0.018$. 
Each line corresponds to the plot for several different values 
of $\mu$ shown in the label.
(Right) $M/M_{\odot}$ versus the star radius $r_s$ for the 
same EOS and the values of $\mu$ as those used in the left panel. 
The central mass density is in the range 
$1 \le \rho_{m0}/\bar{\rho}_m \le 100$. 
\label{fig1}} 
\end{figure}

The iterative solutions around the center of star are given by 
Eqs.~(\ref{Atr=0})-(\ref{Pm}) with $\phi_0=0$. 
Since $\phi(r)=0$ everywhere, we 
integrate Eqs.~(\ref{Ateq}) and (\ref{heq})-(\ref{Peq}) 
from $s=\ln r/\bar{r}=-10$ to the distance $s>40$ (i.e., $r>10^{17}\bar{r}$) 
by using Eqs.~(\ref{Atr=0})-(\ref{Pm}) as the boundary conditions 
around $r=0$. For concreteness we consider the polytropic 
EOS (\ref{yz}) with $\Gamma=5/3$ and $K=0.018$, but  
the qualitative behavior of solutions should be similar for other 
choices of $\Gamma$ and $K$. 
In Fig.~\ref{fig1}, we plot the normalized ADM mass 
$M/M_{\odot}$ versus $\rho_{m0}/\bar{\rho}_m$ (left) 
and $M/M_{\odot}$ versus $r_s$ (right)
for several different values of $\mu$. 
For increasing $\mu$ with a given central matter density $\rho_{m0}$, 
$M$ gets larger. 
This is mostly attributed to the fact that, around the center of star, 
the larger charge density $\rho_{c0}$ leads to the slower 
decrease of $P_m (r)$, according to Eq.~(\ref{Pm}). 
For larger $\mu$, the radius $r_s$ increases due to the extra pressure 
against gravity induced by the Coulomb force. 
This property is confirmed in the right panel of Fig.~\ref{fig1}, 
where the radius for $\mu=0.69$ with a given value of 
$\rho_{m0}$ is much larger than that for $\mu=0$. 

%%%%%%%%%%%%%%%%%%%%%%%%%%%%%%
%%%%%%%%%%%%%%%%%%%%%%%%%%%%%%

As $\mu$ approaches the upper bound (\ref{muup2}), i.e., 
$\mu \simeq 0.7$, the increase of $M$ tends to be significant. 
Even though $M$ is smaller than $M_{\odot}$ for $\mu=0$, 
the presence of a large charge density with $\mu=0.69$ can result in 
the value of $M$ exceeding $10 M_{\odot}$ in the region of low 
matter densities. In the left panel of Fig.~\ref{fig1}, we also find that, 
for $\mu$ close to 0.7, the derivative ${\rm d}M/{\rm d} \rho_{m0}$ 
is negative for most of $\rho_{m0}$
in the range $\bar{\rho}_m \le \rho_{m0} \le 100 \bar{\rho}_m$.
For $\mu>1/\sqrt{2}$, our numerical simulations 
show that gravitationally bounded stars cease to exist. 
Besides the ADM mass $M$, we compute the proper mass $M_p$ 
defined in Eq.~(\ref{Mp}) and find that the binding 
energy $\Delta=M_p-M$ is positive for 
$\mu<1/\sqrt{2}$. As $\mu$ increases toward this upper bound, 
the ADM mass $M$ tends to approach $M_p$ for most of 
$\rho_{m0}$ chosen in Fig.~\ref{fig1}. 
This supports our claim that the gravitationally bounded 
stars can exist for $\mu<1/\sqrt{2}$.
We note that the above results are consistent with 
those obtained in Ref.~\cite{Ray:2003gt} in Einstein-Maxwell theory.

\subsection{Scalarized solutions for $\beta<0$}

\subsubsection{Existence of scalarized solutions}

\begin{figure}[h]
\begin{center}
\includegraphics[height=3.4in,width=3.4in]{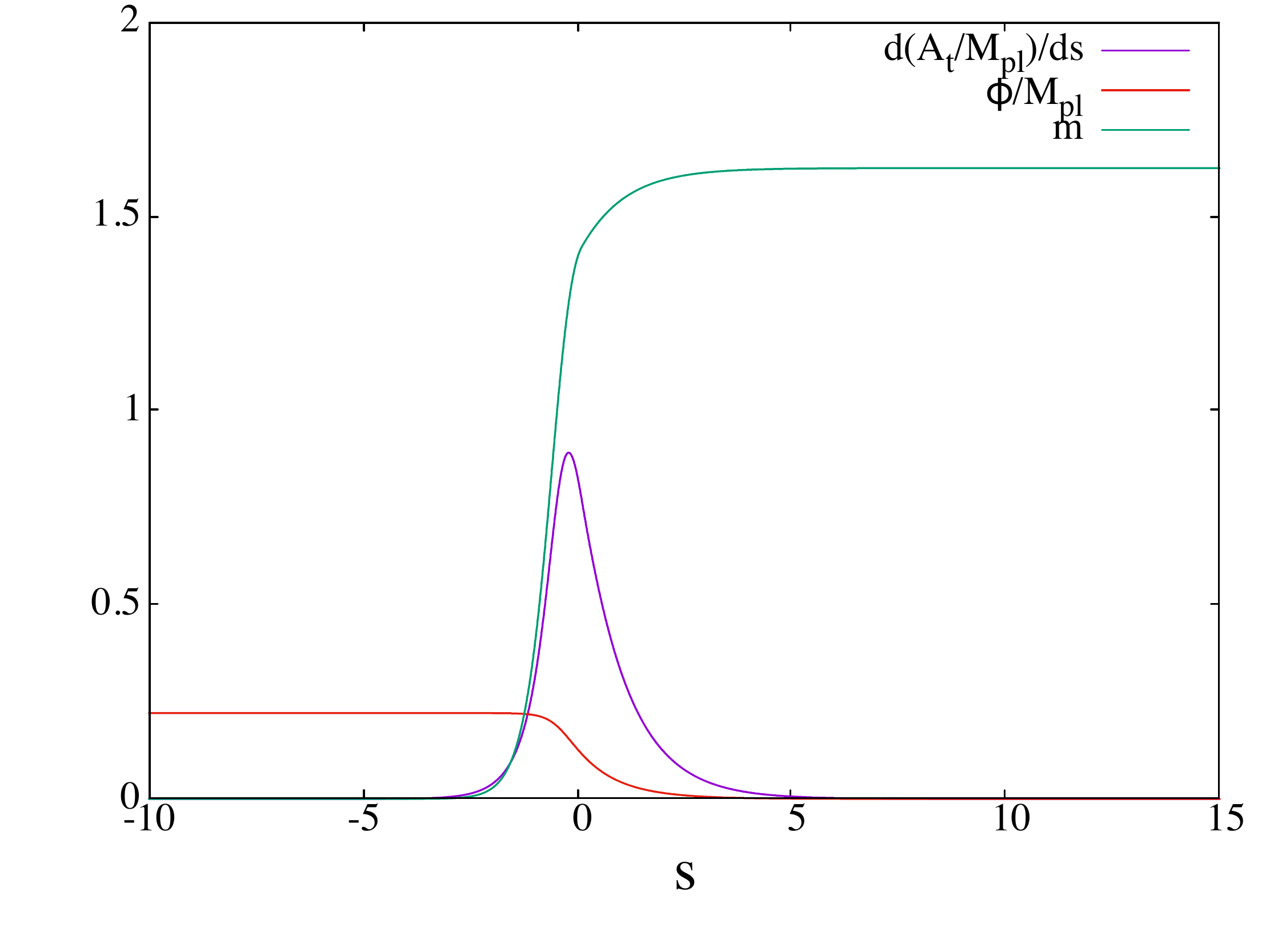}
\includegraphics[height=3.6in,width=3.4in]{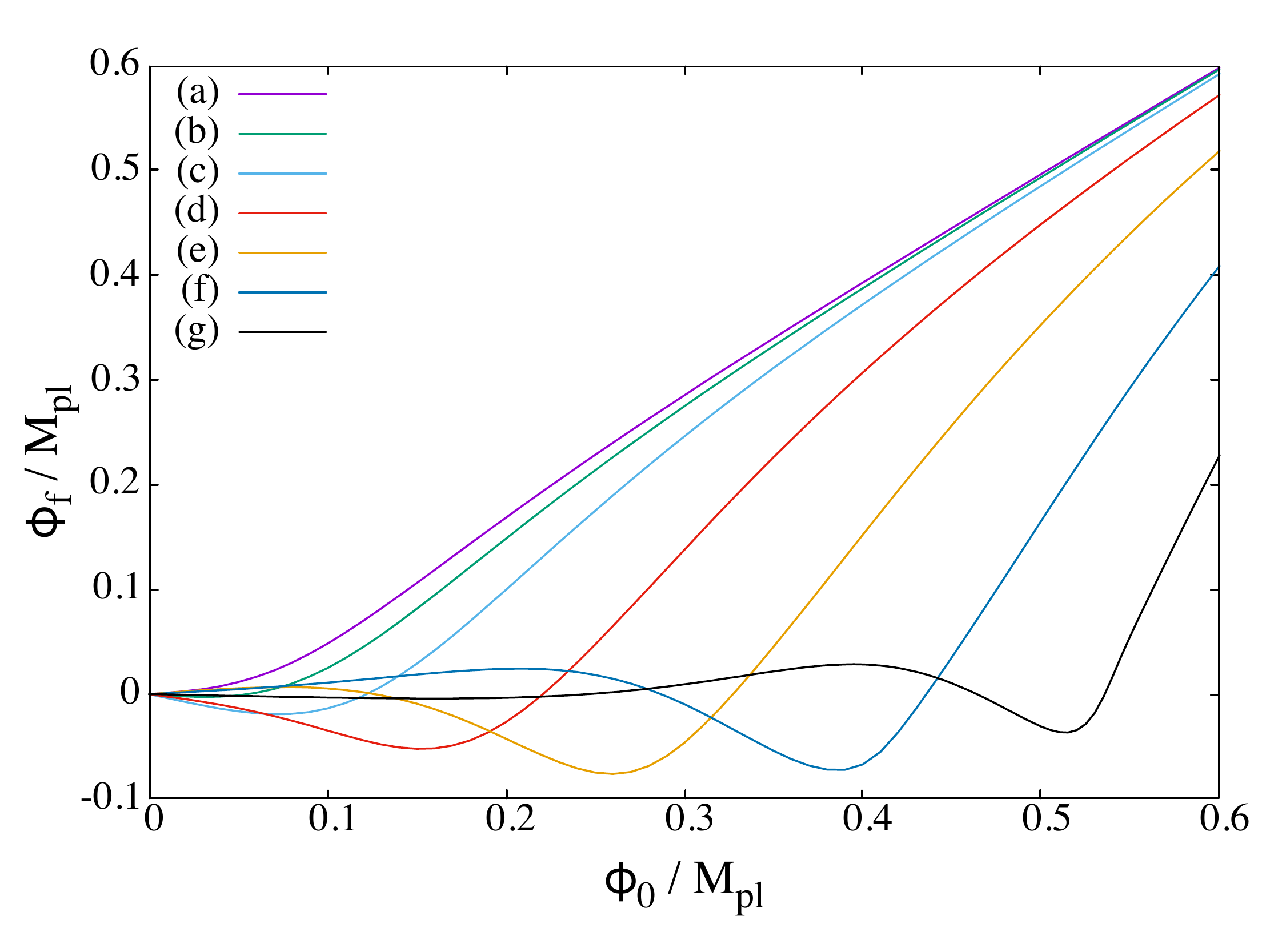}
\end{center}
\caption{(Left) ${\rm d}(A_t/M_{\rm pl})/{\rm d}s$, $\phi/M_{\rm pl}$, and $m$ versus 
$s=\ln r/\bar{r}$ for $\beta=-5$, $\mu=0.68$, and $\chi_0=10$, where $\chi_0$ 
is the value of $\chi$ at $r=0$.
The EOS is the same as that used in Fig.~\ref{fig1}.
The field value $\phi_0 \simeq 0.2208$ at $r=0$ is chosen to satisfy the 
boundary condition $\phi_f=0$.
(Right) The field value $\phi_f$ at spatial infinity normalized by $M_{\rm pl}$
versus $\phi_0/M_{\rm pl}$ for $\beta=-5$ and $\mu=0.68$ 
with the same EOS as in the left panel. Each case corresponds to (a) $\chi_0=1$, 
(b) $\chi_0=10^{1/5}$, (c) $\chi_0=10^{1/2}$, (d) $\chi_0=10$, 
(e) $\chi_0=10^{3/2}$, (f) $\chi_0=10^2$, and (g) $\chi_0=10^{5/2}$.
\label{fig2}} 
\end{figure}

For the negative scalar-gauge coupling ($\beta<0$), we explore the presence 
of scalarized solutions with $\phi(r) \neq 0$ besides the GR branch 
with $\phi(r)=0$. Note that the GR branch corresponds to that 
discussed in Sec.~\ref{nohairsec}. 
We consider a positive value of $\phi_0$ and adopt the polytropic EOS 
same as used in Sec.~\ref{nohairsec}. 
In the left panel of Fig.~\ref{fig2}, we plot an example of the 
scalarized branch for $\beta=-5$ and 
$\mu=0.68$ with $\chi=10$ at $r=0$.
The field value at $r=0$ is determined to be $\phi_0 \simeq 0.2208M_{\rm pl}$
to satisfy the boundary condition $\phi_f \equiv \phi(\infty)=0$
at spatial infinity.
Around $r=0$, the scalar field slowly varies, according to Eq.~(\ref{phir=0}). 
The variation of $\phi(r)$ tends to be significant around the surface of star 
($s \simeq 0.13$). For $r \gg r_s$, the scalar field decreases toward 0 
with its derivative $\phi'(r)$ proportional to $r^{-2}$. 
The derivative of $A_t$ has the dependence ${\rm d}A_t/{\rm d}s \propto r^2$ 
for $r \ll r_s$ and  ${\rm d}A_t/{\rm d}s \propto r^{-1}$ for $r \gg r_s$.
The mass function $m(r)$, which acquires the contributions from $\phi'(r)$ and 
$A_t'(r)$ even outside the star, approaches the constant value 
$m(r) \to 1.62$ as $r \to \infty$.

%%%%%%%%%%%%%%%%%%%%%%%%%%%%%%
%%%%%%%%%%%%%%%%%%%%%%%%%%%%%%

In the right panel of Fig.~\ref{fig2}, we show the dependence of $\phi_f$ on 
the change of $\phi_0$ for seven different central matter densities, 
with $\beta=-5$ and $\mu=0.68$. 
In case (a), there is no nonvanishing value of $\phi_0$ leading to $\phi_f=0$, 
so the GR branch $\phi(r)=0$ is the only allowed solution. 
For $\chi_0 \gtrsim 1.3$, however, there are intersections of the curve 
with $\phi_f=0$ at a point $\phi_0>0$. 
This is the appearance of a scalarized branch where $\phi(r)$ 
changes from $\phi_0$ (at $r=0$) to $0$ (at $r \to \infty$). 
The field profile shown in the left panel of Fig.~\ref{fig2} 
corresponds to case (d) in the right panel, i.e., 
$\phi_0 \simeq 0.2208M_{\rm pl}$ at $\phi_f=0$.

In cases (b)-(d) the shapes of curves in the 
$(\phi_0, \phi_f)$ plane, which are convex downward with the 
existence of the region $\phi_f<0$, are similar to those for 
spontaneous scalarization induced by a nonminimal scalar-field 
coupling with the Ricci scalar \cite{Kase:2020yhw}. 
This shows that, for a wide range of central matter densities, 
the scalar-gauge coupling gives rise to hairy solutions 
which are expected to be
the end points of tachyonic instabilities of 
the GR branch with $\phi(r)=0$.
In order to verify that the new hairy branch is indeed the endpoint
of tachyonic instabilities,
we need perturbative stability analysis \cite{Harada:1998ge}
and numerical simulations \cite{Novak:1998rk}
for the dynamical evolution toward the scalarized solution.

In cases (e) and (f), there are two intersection points of 
the theoretical curve with $\phi_f=0$. 
One of them has the property ${\rm d}\phi_f/{\rm d}\phi_0>0$ around 
$\phi_f=0$, whereas the other has the negative sign of  
${\rm d}\phi_f/{\rm d}\phi_0$. 
The former corresponds to a 0-node solution where $\phi(r)$ 
monotonically decreases toward 0 from $r=0$ to $r \to \infty$.
The latter is known as a 1-node solution where $\phi(r)$ crosses 0 
at finite distance and approaches 0 toward infinity from 
the side $\phi(r)<0$. 
For the 1-node solution with ${\rm d}\phi_f/{\rm d}\phi_0<0$, 
a positive shift of $\phi_0$ leads to 
a negative shift of $\phi_f$, whose property is different from 
the 0-node solution.
Moreover, the 1-node can be regarded as an excited state of solutions, 
while the 0-node should be in a more stable ground state. 
For the computations of $M$ and $r_s$,
we will adopt the values $\phi_0 \simeq 0.3274M_{\rm pl}$ 
and $\phi_0 \simeq 0.4357M_{\rm pl}$ in cases (e) and (f), respectively, 
both of which correspond to the 0-node solutions.

In case (g), the 2-node solution with ${\rm d}\phi_f/{\rm d}\phi_0>0$ 
arises at $\phi_0 \simeq 0.2457M_{\rm pl}$, besides 
the 1-node ($\phi_0 \simeq 0.4639M_{\rm pl}$) and 
the 0-node ($\phi_0 \simeq 0.5356M_{\rm pl}$). 
Since the higher-node solutions should correspond to 
more excited states of the charged star, it is generally expected 
that the 0-node corresponds to 
an endpoint of tachyonic instabilities of the GR branch.
For increasing $\chi_0$ further, the solutions with additional higher nodes 
like the 3-node can appear, but this is the region of extremely
high matter densities which are out of the configuration 
of standard relativistic stars. For this reason, 
we will focus on the matter density in the range 
$\rho_{m0} \le 100\bar{\rho}_m$ (i.e., $\chi_0<68.8$) 
in the following discussion. In this regime, all the scalarized 
field profiles with ${\rm d}\phi_f/{\rm d}\phi_0>0$ at 
$\phi_f=0$ correspond to the 0-node solutions.

\subsubsection{Properties of scalarized solutions}

%%%%%%%%%%%%%%%%%%%%%%%%%%%%%%
\begin{figure}[h]
\begin{center}
\includegraphics[height=3.3in,width=3.4in]{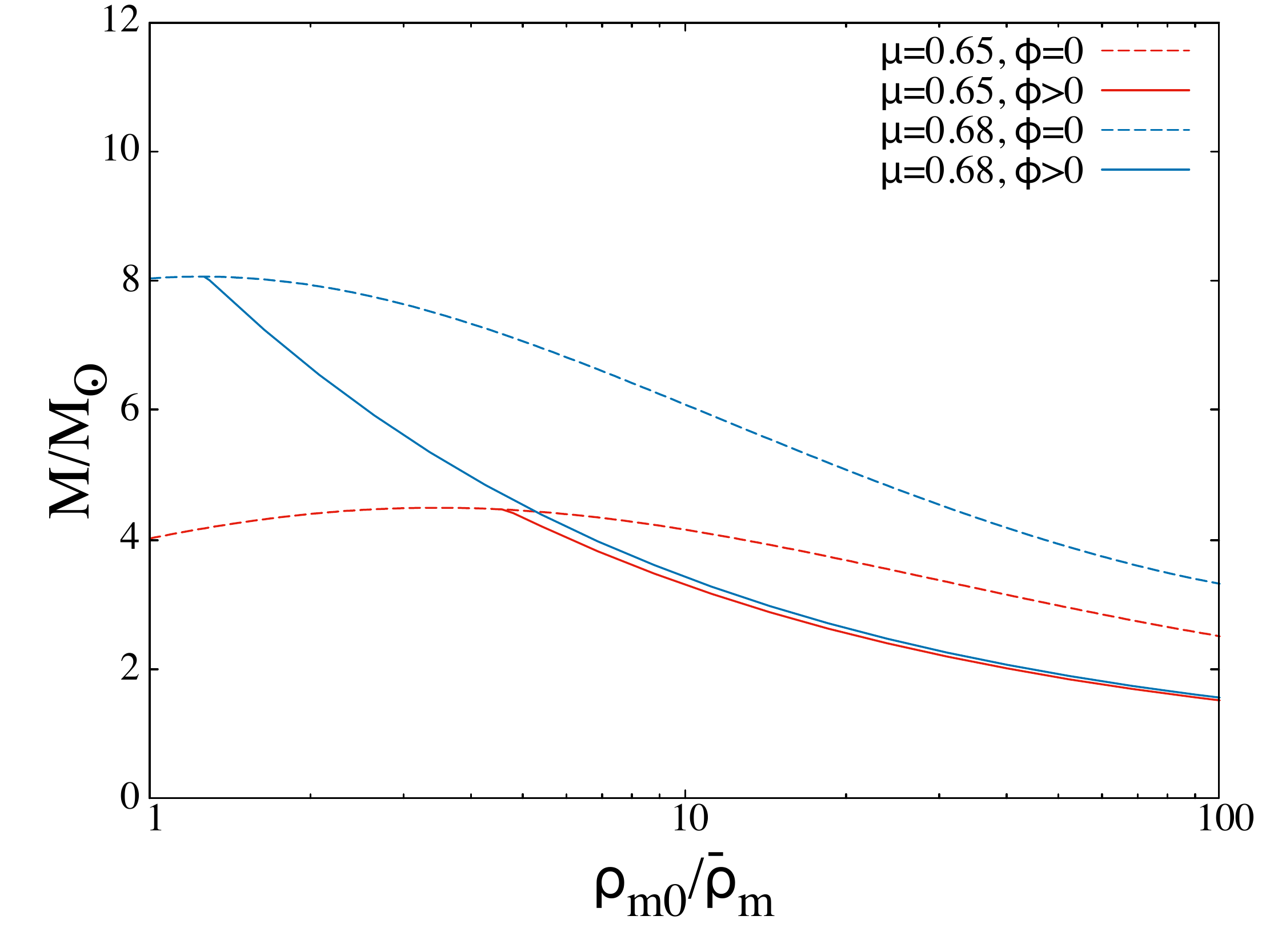}
\includegraphics[height=3.6in,width=3.4in]{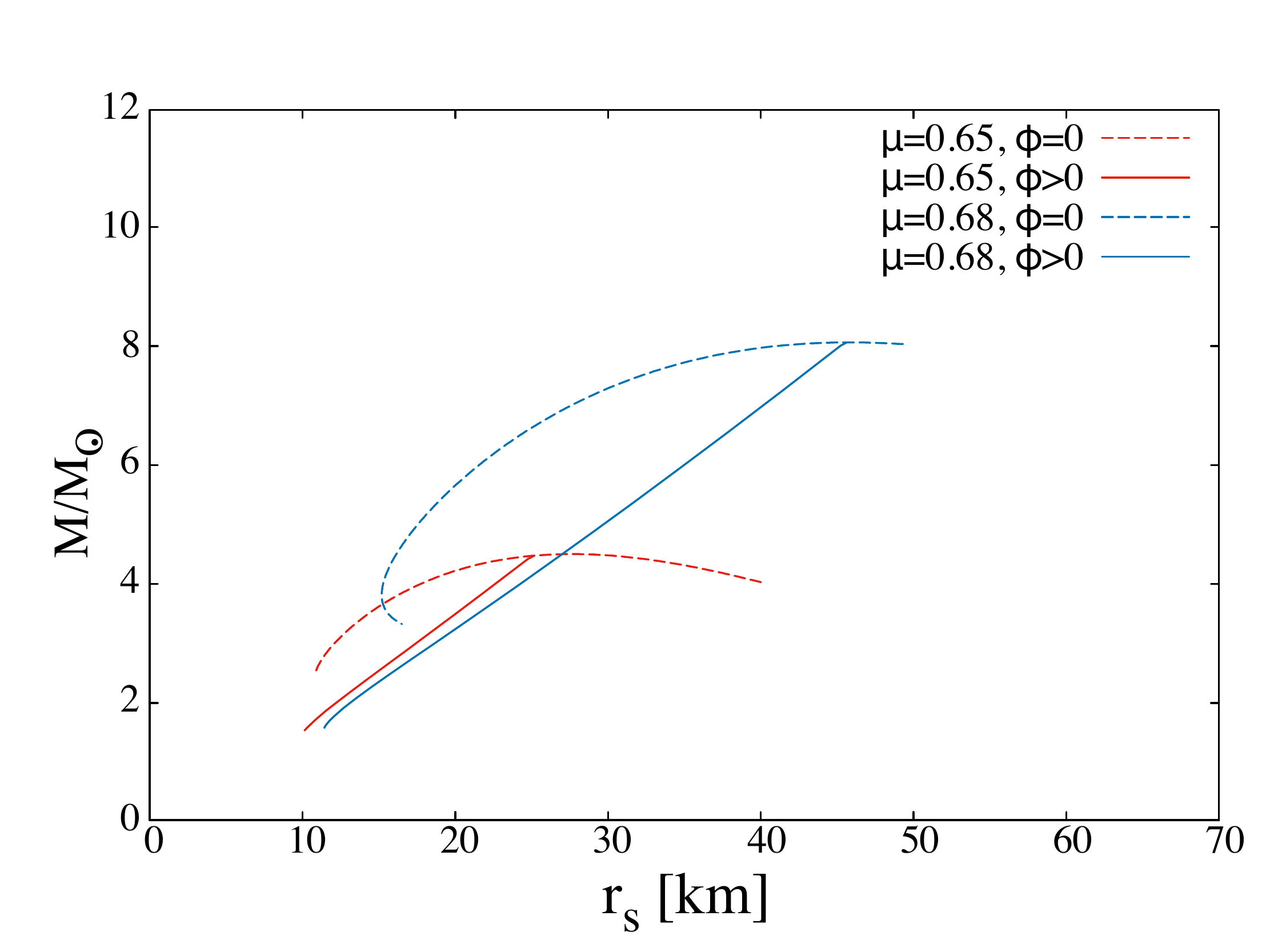}
\end{center}
\caption{$M/M_{\odot}$ versus $\rho_{m0}/\bar{\rho}_m$ (left) 
and $M/M_{\odot}$ versus $r_s$ (right) for $\beta=-5$. 
The red solid and blue solid lines correspond to 
the scalarized branch ($\phi(r) \neq 0$) with $\mu=0.65$ and 
$\mu=0.68$, respectively.
The red dashed and blue dashed lines represent
the GR branch ($\phi(r)=0$) for $\mu=0.65$ and 
$\mu=0.68$, respectively.
The EOS is the same as that used in Fig.~\ref{fig1}. 
In the right panel, the central mass density is 
in the range  
$\bar{\rho}_m \le \rho_{m0} \le 100 \bar{\rho}_m$. 
\label{fig3}} 
\end{figure}
%%%%%%%%%%%%%%%%%%%%%%%%%%%%%%

In Fig.~\ref{fig3}, we plot $M/M_{\odot}$ versus 
$\rho_{m0}/\bar{\rho}_m$ (left) and $M/M_{\odot}$ versus 
$r_s$ (right) with the coupling $\beta=-5$ for two 
different values of $\mu$.
We note that all the nontrivial solutions shown 
in Fig.~\ref{fig3} are the 0-node solutions.
When $\mu=0.68$, the hairy solutions with $\phi(r) \neq 0$
start to appear for the central matter density $\rho_{m0}$ 
larger than $1.3 \bar{\rho}_m$. 
In this regime, the deviations of $M$ and $r_s$ from those 
in the GR branch can be seen in Fig.~\ref{fig3}.
The negative coupling $\beta<0$ with 
$\phi(r) \neq 0$ suppresses the 
Coulomb force $e^{\beta \phi_0^2/M_{\rm pl}^2}\rho_{c0}^2 r^2/6$ 
in Eq.~(\ref{Pm}) around $r=0$. 
The faster decrease of $P_m$ induced by the scalarized branch 
results in smaller values of $r_s$ and $M$ relative to those 
in the GR branch.
Unlike spontaneous scalarization of nonminimally 
coupled scalar-tensor theories \cite{Damour:1993hw,Damour2},
the scalar-gauge coupling gives rise to a charged star configuration 
with the decreased values of $M$.
In the right panel of Fig.~\ref{fig3}, we observe that the ADM mass 
of the scalarized solution is almost proportional to $r_s$.

When $\mu=0.65$ and $\beta=-5$, the scalarized branch starts to 
appear for $\rho_{m0} \gtrsim 4.5 \bar{\rho}_m$.  
The ADM masses of the GR branch exhibit large differences 
between $\mu=0.65$ and $\mu=0.68$, but 
the corresponding masses and radii of scalarized solutions 
are not much different  between these two values of $\mu$.
This can be regarded as the consequence of stabilizations of 
solutions induced by the negative scalar-gauge coupling $\beta$, 
in a way that the slight change of $\mu$ results in only small 
modifications to $M$ and $r_s$.  
We numerically find that, for $\beta=-5$ and $\mu \lesssim 0.5$, 
the scalarized branch does not exist at least in the 
range $\rho_{m0} \le 100\bar{\rho}_m$. 
For $\mu>1/\sqrt{2}$, the negative coupling with $\beta=-5$ 
hardly gives rise to scalarized solutions with $P_m$ decreasing 
toward the surface of star. 
This is related to the fact that the field value $\phi_0$ at $r=0$ 
does not exceed the order of $M_{\rm pl}$, so the Coulomb force 
$e^{\beta \phi_0^2/M_{\rm pl}^2}\rho_{c0}^2 r^2/6$ is not significantly
suppressed compared to that of the GR branch. 
In summary, for $\beta=-5$ and $\rho_{m0} \lesssim 100\bar{\rho}_m$, 
the scalarized charged star solutions exist in the range 
$0.5 \lesssim \mu \lesssim 0.7$.

We have also numerically confirmed the existence of 
0-node scalarized solutions by considering different values of 
$\beta$ and $\mu$. When $\beta=-{\cal O}(0.1)$, we find that 
the scalarized solutions can be present only for
high values of $\rho_{m0}~(\gtrsim {\cal O}(10)\bar{\rho}_m)$ 
with $\mu$ close to $1/\sqrt{2}$. 
For $\beta=-{\cal O}(1)$, 
the scalarized solutions exist for $0.45 \lesssim \mu \lesssim 0.7$ 
in the broader range of $\rho_{m0}$. 
As $\mu$ increases toward the value $1/\sqrt{2}$, 
the ADM mass $M$ is subject to stronger decrease by spontaneous 
scalarization relative to that in the GR branch.
When $|\beta|$ exceeds the order 10, the scalar field does not change much 
inside the star, see Eq.~(\ref{phir=0}). In such cases, there is a tendency that 
the scalarized solutions disappear especially for the values of 
$\mu$ much less than 0.7.

\subsection{Implications for spontanenous scalarization of charged stars}

The results explained above show that, for $\beta=-{\cal O}(1)$ and 
$0.45 \lesssim \mu \lesssim 0.7$, the GR branch can undergo 
spontaneous scalarization toward the nonvanishing 
and nontrivial scalar field solution. 
The masses and radii of the scalarized branch are smaller than those of 
the GR branch as a result of the suppressed Coulomb force.

Without performing detailed numerical simulations about the process 
of spontaneous scalarization, we can speculate to some extent 
how the transition from 
the GR solution to the scalarized branch proceeds.
For $\mu=0.65$ and $\mu=0.68$ shown in Fig.~\ref{fig3},
the 0-node branch is smoothly connected to the GR branch
in low density regimes, showing that the scalarized solution arises 
as the consequence of continuous phase transition from the GR solution.
Fig.~\ref{fig3} also indicates that, for $\beta=-5$, 
the bifurcation point of the 0-node branch from the GR branch
is located in the vicinity of the maximal star mass in GR.
Hence the bifurcation to the 0-node solution occurs around 
the maximum mass of gravitationally bounded charged stars 
in GR.

For a fixed value of $\mu$, the increase of $|\beta|$ within 
$5\lesssim |\beta|\lesssim 10$ shifts bifurcation points 
of the 0-node branch toward the regime of smaller 
densities (larger radii) of the star. 
In particular, for $\mu=0.68$, the bifurcation point is shifted 
to the region $\rho_{m0} \lesssim {\bar \rho}_{m}$.
Thus, spontaneous scalarization to 0-node solutions 
can occur for a wider range of $\rho_{m0}$. 
In the presence of a scalar-gauge coupling, 
tachyonic instabilities excite the scalar field
inside the charged GR star and 
transfer the energies of matter and electric field to the scalar field.
A part of the energy of this system would be radiated away 
via scalar radiation, which reduces the star mass and
makes the 0-node scalarized star perturbatively more stable.

%%%%%%%%%%%%%%%%%%%%%%%%%%%%%%%%%%%%%%%%%%
\section{Conclusions}
\label{concsec}
%%%%%%%%%%%%%%%%%%%%%%%%%%%%%%%%%%%%%%%%%%

In Einstein-Maxwell-scalar theories, we have studied the possibility for the 
occurrence of spontaneous scalarization of charged stars. 
Such theories can be represented by the explicit action (\ref{action}) 
with Eqs.~\eqref{lpf}-\eqref{lc} containing 
the scalar-gauge coupling $-\alpha(\phi)F_{\mu \nu}F^{\mu \nu}/4$ 
and the matter and charge currents as a 
generalization of the Schtuz-Sorkin Lagrangian. 
In Sec.~\ref{basicsec}, we have derived all the field equations of 
motion in the covariant form and shown that the perfect-fluid energy 
momentum tensor $T_{\mu \nu}^{(m)}$ obeys the continuity Eq.~(\ref{Tmunum}) 
with the appearance of a Coulomb force 
on its right hand side.

On the static and spherically symmetric spacetime given 
by the line element (\ref{metric}), 
the background equations of motion are of
the forms (\ref{Ateq})-(\ref{Peq}). 
The field-dependent coupling $\alpha(\phi)$ 
in Eq.~(\ref{phieq}) allows the existence of 
a nonvanishing and nontrivial 
scalar field profile both inside and outside the star. 
This affects the gauge-field derivative $A_t'$ through 
Eq.~(\ref{Ateq}). 
Then, the nontrivial $A_t'$ gives rise to a modified Coulomb force 
which changes the force balance inside the charged star, so that 
its mass $M$ and radius $r_s$ 
are generally subject to modifications.

In Sec.~\ref{sponsec}, we showed that the necessary conditions 
for spontaneous scalarization to occur are characterized 
by Eq.~(\ref{necon}), which leads to the generic form of the coupling 
function \eqref{generic_coupling}.
One of the examples for this realization is the coupling 
$\alpha (\phi)=\exp (-\beta \phi^2/M_{\rm pl}^2)$ with $\beta<0$.
For negative $\beta$, besides the GR branch with $\phi(r)=0$, 
there exists a scalarized branch where the scalar field profile is given by 
Eq.~(\ref{phir=0}) for $r \ll r_s$ and Eq.~(\ref{phiso}) for $r \gg r_s$. 
As the ratio $\mu=M_{\rm pl} \rho_c/\rho_m$ increases 
toward the upper limit $1/\sqrt{2}$, there are significant 
enhancement of $M$ and $r_s$ in the GR branch 
induced by a large Coulomb force.
For $\beta=-{\cal O}(1)$ and $\rho_{m0} \le 100\bar{\rho}_m$, 
we numerically confirmed the existence of 0-node 
scalarized solutions where $\phi(r)$ monotonically decreases from a 
positive value $\phi_0$ (at $r=0$) toward the asymptotic 
value 0 at spatial infinity. 
The scalar-gauge coupling with $\beta<0$ effectively reduces the 
Coulomb force, which results in faster decrease of $P_m (r)$.
As we observe in Fig.~\ref{fig3}, the masses and radii of stars for 
the scaralized branch are smaller than those of the GR branch. 
For this scaralized solution, the modifications to $M$ and $r_s$ induced 
by a small change of $\mu$ around $\mu=0.7$ are much less significant 
in comparison to those of the GR branch.

The appearance of scalarized solutions depends on the central matter 
density $\rho_{m0}$. As $\rho_{m0}$ increases, the scalar-field 
profiles higher than the 0-node start to appear, see the right 
panel of Fig.~\ref{fig2}. 
Since these higher-node solutions correspond to excited states of 
the charged star, we picked up the 0-node solution as a most stable ground 
state of the scalarized branch for the polytropic EOS with $\rho_{m0} \le 100 \bar{\rho}_m$. 
We found that, for $\beta=-{\cal O}(1)$ and $0.45 \lesssim \mu \lesssim 0.7$, 
the scalarized (0-node) solutions can exist for a wide range of $\rho_{m0}$. 
In this regime of $\mu$, the charged GR star with a large Coulomb force 
should undergo spontaneous scalarization toward 
the scalarized branch with smaller values of $M$ and $r_s$. 
This property should also hold for other EOSs because
the Coulomb force $e^{\beta \phi_0^2/M_{\rm pl}^2}\rho_{c0}^2 r^2/6$
appearing in the iterative solution (\ref{Pm}) of $P_m (r)$ 
is generally suppressed for the scalarized branch with negative $\beta$.

It will be of interest to study the stability of our new scalarized solutions  
against perturbations in the odd- and even-parity sectors.
The analysis can be done in the similar way to that performed 
for relativistic stars in nonminimally coupled 
scalar-tensor theories \cite{Kase:2020qvz}. 
We leave this issue for a future work.

%%%%%%%%%%%%%%%%%%
\section*{Acknowledgements}
%%%%%%%%%%%%%%%%%%

MM~was supported by the Portuguese national fund 
through the Funda\c{c}\~{a}o para a Ci\^encia e a Tecnologia
in the scope of the framework of the Decree-Law 57/2016 of August 29
(changed by Law 57/2017 of July 19),
and by the CENTRA through the Project~No.~UIDB/00099/2020.
ST is supported by the Grant-in-Aid for Scientific Research Fund of 
the JSPS No.\,19K03854.

%%%%%%%%%%%%%%%%


\begin{thebibliography}{99}
%%%%%%%%%%%%%%%%

\bibitem{Abbott2016} 
B.~P.~Abbott {\it et al.} [LIGO Scientific and Virgo Collaborations],
%``Observation of Gravitational Waves from a Binary Black Hole Merger,''
Phys.\ Rev.\ Lett.\  {\bf 116}, 061102 (2016)
%doi:10.1103/PhysRevLett.116.061102
[arXiv:1602.03837 [gr-qc]].
%%CITATION = doi:10.1103/PhysRevLett.116.

\bibitem{GW170817} 
B.~P.~Abbott {\it et al.} [LIGO Scientific and Virgo Collaborations],
%``GW170817: Observation of Gravitational Waves from a Binary Neutron Star Inspiral,''
Phys.\ Rev.\ Lett.\  {\bf 119}, 161101 (2017)
%doi:10.1103/PhysRevLett.119.161101
[arXiv:1710.05832 [gr-qc]].

\bibitem{Berti}
E.~Berti {\it et al.},
%``Testing General Relativity with Present and Future Astrophysical Observations,''
Class. Quant. Grav. \textbf{32}, 243001 (2015)
%doi:10.1088/0264-9381/32/24/243001
[arXiv:1501.07274 [gr-qc]].
%642 citations counted in INSPIRE as of 07 Jul 2020

\bibitem{Barack}
L.~Barack {\it et al.},
%``Black holes, gravitational waves and fundamental physics: a roadmap,''
Class. Quant. Grav. \textbf{36}, 143001 (2019)
%doi:10.1088/1361-6382/ab0587
[arXiv:1806.05195 [gr-qc]].

\bibitem{Berti:2018cxi}
E.~Berti, K.~Yagi and N.~Yunes,
%``Extreme Gravity Tests with Gravitational Waves from Compact Binary Coalescences: (I) Inspiral-Merger,''
Gen. Rel. Grav. \textbf{50}, no.4, 46 (2018)
%doi:10.1007/s10714-018-2362-8
[arXiv:1801.03208 [gr-qc]].

\bibitem{Damour:1993hw}
T.~Damour and G.~Esposito-Farese,
%``Nonperturbative strong field effects in tensor - scalar theories of gravitation,''
Phys. Rev. Lett. \textbf{70}, 2220-2223 (1993).
%doi:10.1103/PhysRevLett.70.2220

\bibitem{Damour2} 
T.~Damour and G.~Esposito-Farese,
%``Tensor - scalar gravity and binary pulsar experiments,''
Phys.\ Rev.\ D {\bf 54}, 1474 (1996)
[gr-qc/9602056].

\bibitem{Kanti:1995vq}
P.~Kanti, N.~E.~Mavromatos, J.~Rizos, K.~Tamvakis and E.~Winstanley,
%``Dilatonic black holes in higher curvature string gravity,''
Phys. Rev. D \textbf{54}, 5049-5058 (1996)
%doi:10.1103/PhysRevD.54.5049
[arXiv:hep-th/9511071 [hep-th]].

\bibitem{Alexeev:1996vs}
S.~O.~Alexeev and M.~V.~Pomazanov,
%``Black hole solutions with dilatonic hair in higher curvature gravity,''
Phys. Rev. D \textbf{55}, 2110-2118 (1997)
%doi:10.1103/PhysRevD.55.2110
[arXiv:hep-th/9605106 [hep-th]].

\bibitem{Cooney:2009rr} 
A.~Cooney, S.~DeDeo and D.~Psaltis,
%``Neutron Stars in f(R) Gravity with Perturbative Constraints,''
Phys.\ Rev.\ D {\bf 82}, 064033 (2010)
%doi:10.1103/PhysRevD.82.064033
[arXiv:0910.5480 [astro-ph.HE]].

\bibitem{Arapoglu:2010rz} 
A.~S.~Arapoglu, C.~Deliduman and K.~Y.~Eksi,
%``Constraints on Perturbative f(R) Gravity via Neutron Stars,''
JCAP {\bf 1107}, 020 (2011)
%doi:10.1088/1475-7516/2011/07/020
[arXiv:1003.3179 [gr-qc]].

\bibitem{Rinaldi} 
M.~Rinaldi,
%``Black holes with non-minimal derivative coupling,''
Phys.\ Rev.\ D {\bf 86}, 084048 (2012)
%doi:10.1103/PhysRevD.86.084048
[arXiv:1208.0103 [gr-qc]].

\bibitem{Minami13} 
M.~Minamitsuji,
%``Solutions in the scalar-tensor theory with 
%nonminimal derivative coupling,''
Phys.\ Rev.\ D {\bf 89}, 064017 (2014)
%doi:10.1103/PhysRevD.89.064017
[arXiv:1312.3759 [gr-qc]].

\bibitem{Soti1} 
T.~P.~Sotiriou and S.~Y.~Zhou,
%``Black hole hair in generalized scalar-tensor gravity,''
Phys.\ Rev.\ Lett.\  {\bf 112}, 251102 (2014)
%doi:10.1103/PhysRevLett.112.251102
[arXiv:1312.3622 [gr-qc]].

\bibitem{Soti2} 
T.~P.~Sotiriou and S.~Y.~Zhou,
%``Black hole hair in generalized scalar-tensor 
%gravity: An explicit example,''
Phys.\ Rev.\ D {\bf 90}, 124063 (2014)
%doi:10.1103/PhysRevD.90.124063
[arXiv:1408.1698 [gr-qc]].

\bibitem{Yazadjiev1} 
S.~S.~Yazadjiev, D.~D.~Doneva, K.~D.~Kokkotas and K.~V.~Staykov,
%``Non-perturbative and self-consistent models of neutron 
%stars in R-squared gravity,''
JCAP {\bf 1406}, 003 (2014)
%doi:10.1088/1475-7516/2014/06/003
[arXiv:1402.4469 [gr-qc]].

\bibitem{Babi17} 
E.~Babichev, C.~Charmousis and A.~Leh\'ebel,
%``Asymptotically flat black holes in Horndeski theory and beyond,''
JCAP {\bf 1704}, 027 (2017)
%doi:10.1088/1475-7516/2017/04/027
[arXiv:1702.01938 [gr-qc]].

\bibitem{Babi14} 
E.~Babichev and C.~Charmousis,
%``Dressing a black hole with a time-dependent Galileon,''
JHEP {\bf 1408}, 106 (2014)
%doi:10.1007/JHEP08(2014)106
[arXiv:1312.3204 [gr-qc]].

\bibitem{Koba14} 
T.~Kobayashi and N.~Tanahashi,
%``Exact black hole solutions in shift symmetric scalar\UTF{2013}tensor theories,''
PTEP {\bf 2014}, 073E02 (2014)
%doi:10.1093/ptep/ptu096 
[arXiv:1403.4364 [gr-qc]].

\bibitem{Yazadjiev2} 
S.~S.~Yazadjiev, D.~D.~Doneva and D.~Popchev,
%``Slowly rotating neutron stars in scalar-tensor theories 
%with a massive scalar field,''
Phys.\ Rev.\ D {\bf 93}, 084038 (2016) 
%doi:10.1103/PhysRevD.93.084038
[arXiv:1602.04766 [gr-qc]].

\bibitem{Babi16} 
E.~Babichev, C.~Charmousis, A.~Leh\'ebel and T.~Moskalets,
%``Black holes in a cubic Galileon universe,''
JCAP {\bf 1609}, 011 (2016)
%doi:10.1088/1475-7516/2016/09/011
[arXiv:1605.07438 [gr-qc]].

\bibitem{Tasinato2}
J.~Chagoya, G.~Niz and G.~Tasinato,
%``Black Holes and Neutron Stars in Vector Galileons,''
Class.\ Quant.\ Grav.\  {\bf 34} (2017) no.16,  165002
%doi:10.1088/1361-6382/aa7c01
[arXiv:1703.09555 [gr-qc]].

\bibitem{Minamitsuji}
M.~Minamitsuji,
%``Solutions in the generalized Proca theory with the
% nonminimal coupling to the Einstein tensor,''
Phys.\ Rev.\ D {\bf 94}, 084039 (2016) 
%doi:10.1103/PhysRevD.94.084039 
[arXiv:1607.06278 [gr-qc]].

\bibitem{GPBH}
L.~Heisenberg, R.~Kase, M.~Minamitsuji and S.~Tsujikawa,
%``Hairy black-hole solutions in generalized Proca theories,''
Phys.\ Rev.\ D {\bf 96}, 084049 (2017)
%doi:10.1103/PhysRevD.96.084049
[arXiv:1705.09662 [gr-qc]].

\bibitem{GPBH2} 
L.~Heisenberg, R.~Kase, M.~Minamitsuji and S.~Tsujikawa,
%``Black holes in vector-tensor theories,''
JCAP {\bf 1708}, 024 (2017)
%  doi:10.1088/1475-7516/2017/08/024
[arXiv:1706.05115 [gr-qc]].

\bibitem{Fan} 
Z.~Y.~Fan,
%``Black holes with vector hair,''
JHEP {\bf 1609}, 039 (2016)
%doi:10.1007/JHEP09(2016)039
[arXiv:1606.00684 [hep-th]].

\bibitem{Cisterna} 
A.~Cisterna, M.~Hassaine, J.~Oliva and M.~Rinaldi,
%``Static and rotating solutions for Vector-Galileon theories,''
Phys.\ Rev.\ D {\bf 94}, 104039 (2016).
%doi:10.1103/PhysRevD.94.104039
[arXiv:1609.03430 [gr-qc]].

\bibitem{Babichev17} 
E.~Babichev, C.~Charmousis and M.~Hassaine,
%``Black holes and solitons in an extended Proca theory,''
JHEP {\bf 1705}, 114 (2017)
%  doi:10.1007/JHEP05(2017)114
[arXiv:1703.07676 [gr-qc]].

\bibitem{KMT17} 
R.~Kase, M.~Minamitsuji and S.~Tsujikawa,
%``Relativistic stars in vector-tensor theories,''
Phys.\ Rev.\ D {\bf 97}, 084009 (2018)
%doi:10.1103/PhysRevD.97.084009
[arXiv:1711.08713 [gr-qc]].

\bibitem{Kase:2018owh}
R.~Kase, M.~Minamitsuji and S.~Tsujikawa,
%``Black holes in quartic-order beyond-generalized Proca theories,''
Phys. Lett. B \textbf{782}, 541-550 (2018)
%doi:10.1016/j.physletb.2018.05.078
[arXiv:1803.06335 [gr-qc]].

\bibitem{Kase:2019dqc}
R.~Kase and S.~Tsujikawa,
%``Neutron stars in $f(R)$ gravity and scalar-tensor theories,''
JCAP \textbf{09}, 054 (2019)
%doi:10.1088/1475-7516/2019/09/054
[arXiv:1906.08954 [gr-qc]].

\bibitem{Kobayashi:2018xvr}
T.~Kobayashi and T.~Hiramatsu,
%``Relativistic stars in degenerate higher-order scalar-tensor theories after GW170817,''
Phys. Rev. D \textbf{97}, 104012 (2018).
%doi:10.1103/PhysRevD.97.104012
[arXiv:1803.10510 [gr-qc]].

\bibitem{BenAchour:2018dap}
J.~Ben Achour and H.~Liu,
%``Hairy Schwarzschild-(A)dS black hole solutions in degenerate 
%higher order scalar-tensor theories beyond shift symmetry,''
Phys. Rev. D \textbf{99}, 064042 (2019)  
%doi:10.1103/PhysRevD.99.064042
[arXiv:1811.05369 [gr-qc]].

\bibitem{Motohashi:2019sen}
H.~Motohashi and M.~Minamitsuji,
%``Exact black hole solutions in shift-symmetric quadratic degenerate 
%higher-order scalar-tensor theories,''
Phys. Rev. D \textbf{99}, 064040 (2019).
%doi:10.1103/PhysRevD.99.064040
[arXiv:1901.04658 [gr-qc]].

\bibitem{Minamitsuji:2019tet}
M.~Minamitsuji and J.~Edholm,
%``Black holes with a nonconstant kinetic term in degenerate 
%higher-order scalar tensor theories,''
Phys. Rev. D \textbf{101}, 044034 (2020). 
%doi:10.1103/PhysRevD.101.044034
[arXiv:1912.01744 [gr-qc]].

\bibitem{Harada:1998ge} 
T.~Harada,
%``Neutron stars in scalar tensor theories of gravity 
%and catastrophe theory,''
Phys.\ Rev.\ D {\bf 57}, 4802 (1998)
%doi:10.1103/PhysRevD.57.4802
[gr-qc/9801049].

\bibitem{Novak:1998rk} 
J.~Novak,
%``Neutron star transition to strong scalar field state 
%in tensor scalar gravity,''
Phys.\ Rev.\ D {\bf 58}, 064019 (1998)
% doi:10.1103/PhysRevD.58.064019
[gr-qc/9806022].
%%CITATION = doi:10.1103/PhysRevD.58.064019;%%


\bibitem{Silva:2014fca} 
H.~O.~Silva, C.~F.~B.~Macedo, E.~Berti and L.~C.~B.~Crispino,
%``Slowly rotating anisotropic neutron stars in general relativity 
%and scalar\UTF{2013}tensor theory,''
Class.\ Quant.\ Grav.\  {\bf 32}, 145008 (2015)
% doi:10.1088/0264-9381/32/14/145008
[arXiv:1411.6286 [gr-qc]].

\bibitem{Kase:2020qvz}
R.~Kase, R.~Kimura, S.~Sato and S.~Tsujikawa,
%``Stability of relativistic stars with scalar hairs,''
Phys. Rev. D \textbf{102}, 084037 (2020)
%doi:10.1103/PhysRevD.102.084037
[arXiv:2007.09864 [gr-qc]].

\bibitem{Sotani1}
H.~Sotani and K.~D.~Kokkotas,
%``Probing strong-field scalar-tensor gravity with 
%gravitational wave asteroseismology,''
Phys. Rev. D \textbf{70}, 084026 (2004)
%doi:10.1103/PhysRevD.70.084026
[arXiv:gr-qc/0409066 [gr-qc]].

\bibitem{Freire:2012mg} 
P.~C.~C.~Freire {\it et al.},
%``The relativistic pulsar-white dwarf binary PSR J1738+0333 II. 
%The most stringent test of scalar-tensor gravity,''
Mon.\ Not.\ Roy.\ Astron.\ Soc.\  {\bf 423}, 3328 (2012)
%doi:10.1111/j.1365-2966.2012.21253.x
[arXiv:1205.1450 [astro-ph.GA]].

\bibitem{Sotani2}
H.~Sotani,
%``Scalar gravitational waves from relativistic stars in scalar-tensor gravity,''
Phys. Rev. D \textbf{89}, 064031 (2014)
%doi:10.1103/PhysRevD.89.064031
[arXiv:1402.5699 [astro-ph.HE]].

\bibitem{Pappas:2015npa}
G.~Pappas and T.~P.~Sotiriou,
%``Geodesic properties in terms of multipole moments in scalar\UTF{2013}tensor theories of gravity,''
Mon. Not. Roy. Astron. Soc. \textbf{453}, 2862-2876 (2015)
%doi:10.1093/mnras/stv1819
[arXiv:1505.02882 [gr-qc]].

\bibitem{Sotani:2012eb}
H.~Sotani,
%``Slowly Rotating Relativistic Stars in Scalar-Tensor Gravity,''
Phys. Rev. D \textbf{86}, 124036 (2012)
%doi:10.1103/PhysRevD.86.124036
[arXiv:1211.6986 [astro-ph.HE]].

\bibitem{Doneva:2013qva}
D.~D.~Doneva, S.~S.~Yazadjiev, N.~Stergioulas and K.~D.~Kokkotas,
%``Rapidly rotating neutron stars in scalar-tensor theories of gravity,''
Phys. Rev. D \textbf{88}, 084060 (2013)
%doi:10.1103/PhysRevD.88.084060
[arXiv:1309.0605 [gr-qc]].



\bibitem{Doneva:2014faa}
D.~D.~Doneva, S.~S.~Yazadjiev, K.~V.~Staykov and K.~D.~Kokkotas,
%``Universal I-Q relations for rapidly rotating neutron and 
%strange stars in scalar-tensor theories,''
Phys. Rev. D \textbf{90}, 104021 (2014)
%doi:10.1103/PhysRevD.90.104021
[arXiv:1408.1641 [gr-qc]].

\bibitem{Pani:2014jra}
P.~Pani and E.~Berti,
%``Slowly rotating neutron stars in scalar-tensor theories,''
Phys. Rev. D \textbf{90}, 024025 (2014)
%doi:10.1103/PhysRevD.90.024025
[arXiv:1405.4547 [gr-qc]].

\bibitem{Doneva:2014uma}
D.~D.~Doneva, S.~S.~Yazadjiev, N.~Stergioulas, K.~D.~Kokkotas and T.~M.~Athanasiadis,
%``Orbital and epicyclic frequencies around rapidly rotating 
%compact stars in scalar-tensor theories of gravity,''
Phys. Rev. D \textbf{90}, 044004 (2014)
%doi:10.1103/PhysRevD.90.044004
[arXiv:1405.6976 [astro-ph.HE]].

\bibitem{Minamitsuji:2016hkk}
M.~Minamitsuji and H.~O.~Silva,
%``Relativistic stars in scalar-tensor theories with disformal coupling,''
Phys. Rev. D \textbf{93}, 124041 (2016)
%doi:10.1103/PhysRevD.93.124041
[arXiv:1604.07742 [gr-qc]].

\bibitem{Annulli:2019fzq}
L.~Annulli, V.~Cardoso and L.~Gualtieri,
%``Electromagnetism and hidden vector fields in modified gravity theories: spontaneous and induced vectorization,''
Phys. Rev. D \textbf{99}, 044038 (2019)
%sdoi:10.1103/PhysRevD.99.044038
[arXiv:1901.02461 [gr-qc]].

\bibitem{Ramazanoglu:2019gbz}
F.~M.~Ramazano\u{g}lu,
%``Spontaneous tensorization from curvature coupling and beyond,''
Phys. Rev. D \textbf{99}, 084015 (2019)
%doi:10.1103/PhysRevD.99.084015
[arXiv:1901.10009 [gr-qc]].

\bibitem{Kase:2020yhw}
R.~Kase, M.~Minamitsuji and S.~Tsujikawa,
%``Neutron stars with a generalized Proca hair and spontaneous vectorization,''
Phys. Rev. D \textbf{102}, 024067 (2020)
%doi:10.1103/PhysRevD.102.024067
[arXiv:2001.10701 [gr-qc]].

\bibitem{Minamitsuji:2020pak}
M.~Minamitsuji,
%``Spontaneous vectorization in the presence of vector field coupling 
%to matter,''
Phys. Rev. D \textbf{101}, 104044 (2020)
%doi:10.1103/PhysRevD.101.104044
[arXiv:2003.11885 [gr-qc]].

\bibitem{Ramazanoglu:2017xbl}
F.~M.~Ramazano\u{g}lu,
%``Spontaneous growth of vector fields in gravity,''
Phys. Rev. D \textbf{96}, 064009 (2017).
%doi:10.1103/PhysRevD.96.064009
[arXiv:1706.01056 [gr-qc]].

\bibitem{Ramazanoglu:2018hwk}
F.~M.~Ramazano\u{g}lu,
%``Spontaneous growth of spinor fields in gravity,''
Phys. Rev. D \textbf{98}, 044011(2018)
[erratum: Phys. Rev. D \textbf{100}, 029903 (2019)]
%doi:10.1103/PhysRevD.98.044011
[arXiv:1804.00594 [gr-qc]].

\bibitem{Minamitsuji:2020hpl}
M.~Minamitsuji,
%``Stealth spontaneous spinorization of relativistic stars,''
Phys. Rev. D \textbf{102}, 044048 (2020) 
%doi:10.1103/PhysRevD.102.044048
[arXiv:2008.12758 [gr-qc]].

\bibitem{Doneva:2017bvd}
D.~D.~Doneva and S.~S.~Yazadjiev,
%``New Gauss-Bonnet Black Holes with Curvature-Induced Scalarization 
%in Extended Scalar-Tensor Theories,''
Phys. Rev. Lett. \textbf{120}, 131103 (2018)
%doi:10.1103/PhysRevLett.120.131103
[arXiv:1711.01187 [gr-qc]].

\bibitem{Silva:2017uqg} 
H.~O.~Silva, J.~Sakstein, L.~Gualtieri, T.~P.~Sotiriou and E.~Berti,
%``Spontaneous scalarization of black holes and compact stars 
%from a Gauss-Bonnet coupling,''
Phys.\ Rev.\ Lett.\  {\bf 120}, 131104 (2018)
%doi:10.1103/PhysRevLett.120.131104
[arXiv:1711.02080 [gr-qc]].

\bibitem{Silva:2018qhn}
H.~O.~Silva, C.~F.~B.~Macedo, T.~P.~Sotiriou, L.~Gualtieri, J.~Sakstein and E.~Berti,
%``Stability of scalarized black hole solutions in scalar-Gauss-Bonnet gravity,''
Phys. Rev. D \textbf{99}, 064011 (2019)
%doi:10.1103/PhysRevD.99.064011
[arXiv:1812.05590 [gr-qc]].

\bibitem{Antoniou:2017acq} 
G.~Antoniou, A.~Bakopoulos and P.~Kanti,
%``Evasion of No-Hair Theorems and Novel Black-Hole 
%Solutions in Gauss-Bonnet Theories,''
Phys.\ Rev.\ Lett.\  {\bf 120},  131102 (2018)
%doi:10.1103/PhysRevLett.120.131102
[arXiv:1711.03390 [hep-th]].

\bibitem{Antoniou:2017hxj} 
G.~Antoniou, A.~Bakopoulos and P.~Kanti,
%``Black-Hole Solutions with Scalar Hair in Einstein-Scalar-Gauss-Bonnet Theories,''
Phys.\ Rev.\ D {\bf 97}, 084037 (2018)
%doi:10.1103/PhysRevD.97.084037
[arXiv:1711.07431 [hep-th]].

\bibitem{Minamitsuji:2018xde} 
M.~Minamitsuji and T.~Ikeda,
%``Scalarized black holes in the presence of the 
%coupling to Gauss-Bonnet gravity,''
Phys.\ Rev.\ D {\bf 99}, 044017 (2019)
%doi:10.1103/PhysRevD.99.044017
[arXiv:1812.03551 [gr-qc]].

\bibitem{Cunha} 
P.~V.~P.~Cunha, C.~A.~R.~Herdeiro and E.~Radu,
%``Spontaneously Scalarized Kerr Black Holes in Extended 
%Scalar-Tensor\UTF{2013}Gauss-Bonnet Gravity,''
Phys.\ Rev.\ Lett.\  {\bf 123}, 011101 (2019)
[arXiv:1904.09997 [gr-qc]].

\bibitem{Dima:2020yac}
A.~Dima, E.~Barausse, N.~Franchini and T.~P.~Sotiriou,
%``Spin-induced black hole spontaneous scalarization,''
Phys. Rev. Lett. \textbf{125}, 231101 (2020)
%doi:10.1103/PhysRevLett.125.231101
[arXiv:2006.03095 [gr-qc]].

\bibitem{Hod:2020jjy}
S.~Hod,
%``Onset of spontaneous scalarization in spinning Gauss-Bonnet black holes,''
Phys. Rev. D \textbf{102}, no.8, 084060 (2020)
%doi:10.1103/PhysRevD.102.084060
[arXiv:2006.09399 [gr-qc]].

\bibitem{Herdeiro:2020wei}
C.~A.~R.~Herdeiro, E.~Radu, H.~O.~Silva, T.~P.~Sotiriou and N.~Yunes,
%``Spin-induced scalarized black holes,''
Phys. Rev. Lett. \textbf{126}, 011103 (2021)
%doi:10.1103/PhysRevLett.126.011103
[arXiv:2009.03904 [gr-qc]].

\bibitem{Berti:2020kgk}
E.~Berti, L.~G.~Collodel, B.~Kleihaus and J.~Kunz,
%``Spin-induced black-hole scalarization in 
%Einstein-scalar-Gauss-Bonnet theory,''
Phys. Rev. Lett. \textbf{126}, 011104 (2021)
%doi:10.1103/PhysRevLett.126.011104
[arXiv:2009.03905 [gr-qc]].

\bibitem{Doneva:2021dqn}
D.~D.~Doneva and S.~S.~Yazadjiev,
%``Dynamics of the nonrotating and rotating black hole scalarization,''
Phys. Rev. D \textbf{103} (2021) no.6, 064024
%doi:10.1103/PhysRevD.103.064024
[arXiv:2101.03514 [gr-qc]].

\bibitem{Gao:2018acg}
Y.~X.~Gao, Y.~Huang and D.~J.~Liu,
%``Scalar perturbations on the background of Kerr black holes 
%in the quadratic dynamical Chern-Simons gravity,''
Phys. Rev. D \textbf{99}, 044020 (2019)
%doi:10.1103/PhysRevD.99.044020
[arXiv:1808.01433 [gr-qc]].

\bibitem{Myung:2020etf}
Y.~S.~Myung and D.~C.~Zou,
%``Onset of rotating scalarized black holes 
%in Einstein-Chern-Simons-Scalar theory,''
Phys. Lett. B \textbf{814}, 136081 (2021) 
%doi:10.1016/j.physletb.2021.136081
[arXiv:2012.02375 [gr-qc]].

\bibitem{Doneva:2021dcc}
D.~D.~Doneva and S.~S.~Yazadjiev,
%``Spontaneously scalarized black holes in dynamical Chern-Simons 
%gravity: dynamics and equilibrium solutions,''
Phys. Rev. D \textbf{103}, 083007 (2021)
%doi:10.1103/PhysRevD.103.083007
[arXiv:2102.03940 [gr-qc]].

\bibitem{Stefanov} 
I.~Z.~Stefanov, S.~S.~Yazadjiev and M.~D.~Todorov,
%``Phases of 4D scalar-tensor black holes coupled to 
%Born-Infeld nonlinear electrodynamics,''
Mod.\ Phys.\ Lett.\ A {\bf 23}, 2915 (2008)
[arXiv:0708.4141 [gr-qc]].

\bibitem{Herdeiro1} 
C.~A.~R.~Herdeiro, E.~Radu, N.~Sanchis-Gual and J.~A.~Font,
%``Spontaneous Scalarization of Charged Black Holes,''
Phys.\ Rev.\ Lett.\  {\bf 121}, 101102 (2018)
[arXiv:1806.05190 [gr-qc]].

\bibitem{Myung:2018vug}
Y.~S.~Myung and D.~C.~Zou,
%``Instability of Reissner\textendash{}Nordstr\"om black hole 
%in Einstein-Maxwell-scalar theory,''
Eur. Phys. J. C \textbf{79}, no.3, 273 (2019)
%doi:10.1140/epjc/s10052-019-6792-6
[arXiv:1808.02609 [gr-qc]].

\bibitem{Herdeiro2} 
P.~G.~S.~Fernandes, C.~A.~R.~Herdeiro, A.~M.~Pombo, E.~Radu and 
N.~Sanchis-Gual,
%``Spontaneous Scalarisation of Charged Black Holes: 
%Coupling Dependence and Dynamical Features,''
Class.\ Quant.\ Grav.\  {\bf 36}, 134002 (2019)
[arXiv:1902.05079 [gr-qc]].

\bibitem{Brihaye:2019kvj}
Y.~Brihaye and B.~Hartmann,
%``Spontaneous scalarization of charged black holes at the 
%approach to extremality,''
Phys. Lett. B \textbf{792}, 244-250 (2019)
%doi:10.1016/j.physletb.2019.03.043
[arXiv:1902.05760 [gr-qc]].

\bibitem{Myung:2019oua}
Y.~S.~Myung and D.~C.~Zou,
%``Stability of scalarized charged black holes in the Einstein\textendash{}
%Maxwell\textendash{}Scalar theory,''
Eur. Phys. J. C \textbf{79}, no.8, 641 (2019)
%doi:10.1140/epjc/s10052-019-7176-7
[arXiv:1904.09864 [gr-qc]].

\bibitem{Herdeiro3} 
P.~G.~S.~Fernandes, C.~A.~R.~Herdeiro, A.~M.~Pombo, E.~Radu 
and N.~Sanchis-Gual,
%``Charged black holes with axionic-type couplings: Classes of solutions 
%and dynamical scalarization,''
Phys.\ Rev.\ D {\bf 100}, 084045 (2019)
[arXiv:1908.00037 [gr-qc]].

\bibitem{Ikeda:2019okp}
T.~Ikeda, T.~Nakamura and M.~Minamitsuji,
%``Spontaneous scalarization of charged black holes in the 
%Scalar-Vector-Tensor theory,''
Phys. Rev. D \textbf{100}, 104014 (2019)
%doi:10.1103/PhysRevD.100.104014
[arXiv:1908.09394 [gr-qc]].

\bibitem{Hod:2020ljo}
S.~Hod,
%``Spontaneous scalarization of charged Reissner-Nordstr\"om 
%black holes: Analytic treatment along the existence line,''
Phys. Lett. B \textbf{798}, 135025 (2019)
%doi:10.1016/j.physletb.2019.135025
[arXiv:2002.01948 [gr-qc]].

\bibitem{Gibbons:1987ps}
G.~W.~Gibbons and K.~i.~Maeda,
%``Black Holes and Membranes in Higher Dimensional Theories 
%with Dilaton Fields,''
Nucl. Phys. B \textbf{298}, 741-775 (1988).

\bibitem{Garfinkle:1990qj}
D.~Garfinkle, G.~T.~Horowitz and A.~Strominger,
%``Charged black holes in string theory,''
Phys. Rev. D \textbf{43}, 3140 (1991)
[erratum: Phys. Rev. D \textbf{45}, 3888 (1992)].

\bibitem{Mignemi:1992nt}
S.~Mignemi and N.~R.~Stewart,
%``Charged black holes in effective string theory,''
Phys. Rev. D \textbf{47}, 5259-5269 (1993)
%doi:10.1103/PhysRevD.47.5259
[arXiv:hep-th/9212146 [hep-th]].

\bibitem{Torii:1996yi}
T.~Torii, H.~Yajima and K.~i.~Maeda,
%``Dilatonic black holes with Gauss-Bonnet term,''
Phys. Rev. D \textbf{55}, 739-753 (1997)
%doi:10.1103/PhysRevD.55.739
[arXiv:gr-qc/9606034 [gr-qc]].

\bibitem{Herdeiro:2020htm}
C.~A.~R.~Herdeiro, T.~Ikeda, M.~Minamitsuji, T.~Nakamura and E.~Radu,
%``Spontaneous scalarization of a conducting sphere in Maxwell-scalar models,''
Phys. Rev. D \textbf{103}, no.4, 044019 (2021)
%doi:10.1103/PhysRevD.103.044019
[arXiv:2009.06971 [gr-qc]].

\bibitem{Ray:2003gt}
S.~Ray, A.~L.~Espindola, M.~Malheiro, J.~P.~S.~Lemos and V.~T.~Zanchin,
%``Electrically charged compact stars and formation of charged black holes,''
Phys. Rev. D \textbf{68}, 084004 (2003)
%doi:10.1103/PhysRevD.68.084004
[arXiv:astro-ph/0307262 [astro-ph]].

\bibitem{Bekenstein:1971ej}
J.~D.~Bekenstein,
%``Hydrostatic Equilibrium and Gravitational Collapse 
%of Relativistic Charged Fluid Balls,''
Phys. Rev. D \textbf{4}, 2185-2190 (1971).
%doi:10.1103/PhysRevD.4.2185

\bibitem{deFelice:1999qp}
F.~de Felice, S.~m.~Liu and Y.~q.~Yu,
%``Relativistic charged spheres. 2. Regularity and stability,''
Class. Quant. Grav. \textbf{16}, 2669-2680 (1999)
%doi:10.1088/0264-9381/16/8/307
[arXiv:gr-qc/9905099 [gr-qc]].

\bibitem{Anninos:2001yb}
P.~Anninos and T.~Rothman,
%``Instability of extremal relativistic charged spheres,''
Phys. Rev. D \textbf{65}, 024003 (2002)
%doi:10.1103/PhysRevD.65.024003
[arXiv:gr-qc/0108082 [gr-qc]].

\bibitem{Sorkin}
B.~F.~Schutz and R.~Sorkin,
%``Variational aspects of relativistic field theories,
%with application to perfect fluids,''
Annals Phys.\  {\bf 107}, 1 (1977).

\bibitem{Brown} 
J.~D.~Brown,
%``Action functionals for relativistic perfect fluids,''
Class.\ Quant.\ Grav.\  {\bf 10}, 1579 (1993)
%doi:10.1088/0264-9381/10/8/017
[gr-qc/9304026].

\bibitem{DGS}
A.~De Felice, J.~M.~Gerard and T.~Suyama,
%``Cosmological perturbations of a perfect fluid and 
%noncommutative variables,''
Phys.\ Rev.\ D {\bf 81}, 063527 (2010)
%doi:10.1103/PhysRevD.81.063527
[arXiv:0908.3439 [gr-qc]].

\bibitem{Kase:2020hst}
R.~Kase and S.~Tsujikawa,
%``General formulation of cosmological perturbations in scalar-tensor 
%dark energy coupled to dark matter,''
JCAP \textbf{11}, 032 (2020)
%doi:10.1088/1475-7516/2020/11/032
[arXiv:2005.13809 [gr-qc]].

\bibitem{Tsujikawa:2020die}
R.~Kase and S.~Tsujikawa,
%``Instability of compact stars with a nonminimal scalar-derivative coupling,''
JCAP \textbf{01}, 008 (2021)
%doi:10.1088/1475-7516/2021/01/008
[arXiv:2008.13350 [gr-qc]].

\end{thebibliography}
\end{document}